\documentclass[fleqn,usenatbib]{mnras}

\usepackage{newtxtext,newtxmath}

\usepackage[T1]{fontenc}

\DeclareRobustCommand{\VAN}[3]{#2}
\let\VANthebibliography\thebibliography
\def\thebibliography{\DeclareRobustCommand{\VAN}[3]{##3}\VANthebibliography}

\usepackage{graphicx}	
\usepackage{amsmath}	
\usepackage{threeparttable}
\usepackage{hyperref}
\usepackage{array}
\usepackage{dsnr}
\usepackage{subcaption}

\newcommand{\vfifty}{\ifmmode v_{50\%}\else $v_{50\%}$\fi}
\newcommand{\vfiftyave}{\ifmmode \langle v_{50\%}\rangle \else $\langle
  v_{50\%}\rangle$\fi}
\newcommand{\vtsigr}{\ifmmode v_{02\%}\else $v_{02\%}$\fi}
\newcommand{\vtsig}{\ifmmode v_{98\%}\else $v_{98\%}$\fi}
\newcommand{\vtsigave}{\ifmmode \langle v_{98\%}\rangle \else $\langle
  v_{98\%}\rangle$\fi}
\newcommand{\weq}{\ifmmode W_\mathrm{eq}\else $W_\mathrm{eq}$\fi}
\newcommand{\weqa}{\ifmmode W_\mathrm{eq}^\mathrm{abs}\else $W_\mathrm{eq}^\mathrm{abs}$\fi}
\newcommand{\weqe}{\ifmmode W_\mathrm{eq}^\mathrm{em}\else $W_\mathrm{eq}^\mathrm{em}$\fi}
\newcommand{\ebvg}{\ifmmode \ebv_\mathrm{gas}\else \ebv$_\mathrm{gas}$\fi}
\newcommand{\ebvs}{\ifmmode \ebv_\star\else \ebv$_\star$\fi}
\newcommand{\RNum}[1]{\uppercase\expandafter{\romannumeral #1\relax}}
\renewcommand\ion[2]{#1$\;${%
\ifx\@currsize\normalsize\small \else
\ifx\@currsize\small\footnotesize \else
\ifx\@currsize\footnotesize\scriptsize \else
\ifx\@currsize\scriptsize\tiny \else
\ifx\@currsize\large\normalsize \else
\ifx\@currsize\Large\large
\fi\fi\fi\fi\fi\fi
\rmfamily\RNum{#2}}\relax}%
\renewcommand\sun{\ifmmode \odot\else \hbox{$\odot$}\fi}

\title[Dusty, Neutral Flows]{The spatially-resolved gas and dust connection in neutral inflows and outflows in nearby AGN}

\author[D. S. N. Rupke et al.]{David S. N. Rupke,$^{1,2}$\thanks{E-mail: \href{mailto:drupke@gmail.com}{drupke@gmail.com} (DSNR)}
Adam D. Thomas,$^{2}$
Michael A. Dopita$^{2}$\footnotemark[2]
\\
$^{1}$Department of Physics, Rhodes College, Memphis, TN 38112, USA; \\
$^{2}$Research School of Astronomy \& Astrophysics, Mount Stromlo Observatory, Weston Creek, ACT 2611, Australia\\
}

\date{Accepted 2021 March 5. Received 2021 February 9; in original form 2020 July 13}

\pubyear{2021}

\begin{document}
\label{firstpage}
\pagerange{\pageref{firstpage}--\pageref{lastpage}}
\maketitle

\begin{abstract}
Dusty, neutral outflows and inflows are a common feature of nearby star-forming galaxies. We characterize these flows in eight galaxies---mostly AGN---selected for their widespread \nad\ signatures from the Siding Spring Southern Seyfert Spectroscopic Snapshot Survey (S7). This survey employs deep, wide field-of-view integral field spectroscopy at moderate spectral resolution ($R=7000$ at \nad). We significantly expand the sample of sightlines in external galaxies in which the spatially-resolved relationship has been studied between cool, neutral gas properties---$N$(\ion{Na}{1}), \weq(\nad)---and dust---\ebv\ from both stars and gas. Our sample shows strong, significant correlations of total \weq\ with \ebvs\ and $g-i$ colour within individual galaxies; correlations with \ebvg\ are present but weaker. Regressions yield slope variations from galaxy to galaxy and intrinsic scatter $\sim$1~\AA. The sample occupies regions in the space of $N$(\ion{Na}{1}) and \weqa\ vs. \ebv$_\mathrm{gas}$ that are consistent with extrapolations from other studies to higher colour excess [$\ebv_\mathrm{gas}\sim1$)]. For perhaps the first time in external galaxies, we detect inverse P Cygni profiles in the \nad\ line, presumably due to inflowing gas. Via Doppler shifted \nad\ absorption and emission lines,  we find ubiquitous flows that differ from stellar rotation by $\ga$100~\kms\ or have $|v_{abs} - v_{em}|\ga100$~\kms. Inflows and outflows extend toward the edge of the detected stellar disk/FOV, together subtend 10-40\%\ of the projected disk, and have similar mean $N$(\ion{Na}{1}) and \weq(\nad). Outflows are consistent with minor-axis or jet-driven flows, while inflows tend toward the projected major axis. The inflows may result from non-axisymmetric potentials, tidal motions, or halo infall.
\end{abstract}

\begin{keywords}
galaxies: Seyfert -- galaxies: ISM -- galaxies: kinematics and dynamics -- ISM: dust, extinction
\end{keywords}



\section{INTRODUCTION} \label{sec:introduction}

Deep, wide-field integral field spectroscopy (IFS) of individual
galaxies is now widespread due to the deployment of sensitive
instruments on telescopes of all sizes. A few existing surveys of galaxies with
wide-field IFSs (fields of view, or FOVs, of 20\arcsec--60\arcsec) now have sample sizes $>$10$^2$ \citep[see a summary in Appendix~A of ][]{2020ARA&A..58...99S}. This cohort of surveys includes the completed Siding Spring Southern Seyfert Spectroscopic Snapshot Survey (S7; \citealt{2015ApJS..217...12D,2017ApJS..232...11T}.) The S7 data were
collected with the Wide Field Spectrograph (WiFeS;
\citealt{2007Ap&SS.310..255D,2010Ap&SS.327..245D}), which has moderate
spectral resolution ($R = 7000$ and 3000 in the red and blue,
respectively) and broad wavelength coverage (350--700~nm) over a FOV of
25\arcsec$\times$38\arcsec. WiFeS is thus well-suited to line and kinematic studies of nearby galaxies, which fill a substantial portion of its FOV and are well-matched to its sensitivity (20--35\% throughput) and
spatial resolution (limited by the modest seeing at Siding Spring and sampled by 1\arcsec$\times$1\arcsec\ spaxels).

\renewcommand{\thefootnote}{\mbox{{$\fnsymbol{footnote}$}}}
\footnotetext[2]{Deceased}
\renewcommand{\thefootnote}{\arabic{footnote}}

S7 uniquely targets a large sample of AGN at velocity resolution
useful for probing non-circular motions (43/100~\kms\ in the
red/blue). These non-circular motions---inflows feeding star formation and the accretion disk and outflows driven by accretion energy---are a primary focus for understanding the evolution of nuclear supermassive black
holes. 3D data like IFS is the optimal tool for detecting and
characterizing these motions because of its capacity to disentangle
these inflow and outflow motions from, e.g., simple rotation
\citep[e.g.,][]{2013ApJ...768...75R}. Other IFS surveys of nearby AGN with wide-field IFSs are in progress: the CARS, MAGNUM, and MURALES surveys of 41, 73, and 37 nearby AGN, respectively \citep{2017Msngr.169...42H,2017FrASS...4...46V,2019A&A...632A.124B}.

One of the most useful optical tracers of outflows and inflows in
nearby galaxies is the \nadl\ resonant line. It has been used
successfully to detect and parameterize ubiquitous cool, neutral
outflows in dusty star-forming and luminous active galaxies at
$z \la 0.5$
\citep[e.g.,][]{2000ApJS..129..493H,2005ApJS..160...87R,2005ApJS..160..115R,2005ApJ...632..751R}. Long-slit
and IFS observations of small samples
\citep{2005ApJ...621..227M,2010ApJ...724.1430S,2011ApJ...729L..27R,2013ApJ...768...75R,2015ApJ...801..126R,2017ApJ...850...40R,2019AA...623A.171P}
show that these massive outflows are typically collimated by
kiloparsec-scale disks but often extend to radii 10~kpc or
greater. Stacking Sloan Digital Sky Survey (SDSS) and Mapping Nearby
Galaxies at APO (MaNGA) spectra has confirmed that these
outflows are dusty, common, and large-scale (out to R$_e$) even in
normal star-forming galaxies
\citep{2010AJ....140..445C,2019MNRAS.482.4111R,2019AA...622A.188C,2020MNRAS.493.3081R}. These studies mostly rely on blueshifted resonant absorption to characterize
outflows, but also find that systemic or redshifted emission is
common. The stacked, spatially-integrated emission is most prominent
at the lowest values of galaxy inclination $i$, star formation rate
(SFR) or star formation rate surface density ($\Sigma_\mathrm{SFR}$),
mass, and extinction. Similarly, redshifted absorption has shown that
inflows dominate at $i\ga50^\circ$ \citep{2019MNRAS.482.4111R}.

Unlike surveys of star-forming galaxies, \nad\ studies of cool,
neutral gas flows in active galactic nuclei (AGN) have produced
conflicting results. Some single-aperture \nad\ studies of low-$z$ AGN
find no evidence for \nad\ outflows
\citep{2014MNRAS.440.3202V,2016MNRAS.456L..25S,2017AA...603A..99P}. Others
find that these outflows are present, but at similar rates and with
similar properties to those in galaxies without a detected AGN
\citep{2010ApJ...708.1145K,2019MNRAS.482.4111R,2019MNRAS.486.1608N},
suggesting that the AGN is not energetically important. However, most
of these studies target AGN with luminosities below the quasar
threshold ($L_\mathrm{bol}~\sim10^{45}$~erg~s$^{-1}$), and at higher
luminosities the neutral outflow properties of galaxies may scale with
AGN or black hole properties
\citep{2013ApJ...768...75R,2017ApJ...850...40R}, similar to the
molecular phase
\citep[e.g.,][]{2014A&A...562A..21C,2020A&A...633A.134L}.

\begin{table*}
    \centering
    \caption{Sample properties. Heliocentric redshifts are derived from stellar velocity maps in the current study (Section~\ref{sec:data}); 1$\sigma$ errors are $\delta z = 0.00002$.. Spectral types \citep{2015ApJS..217...12D,2017ApJS..232...11T} are SB $=$ starburst,
    PSB $=$ post-starburst, and Sy2 $=$ Seyfert 2. $L(\oth)$ is the de-reddened nuclear \othl\ luminosity, expressed as log($L$/erg~s$^{-1}$) \citep{2015ApJS..217...12D,2017ApJS..232...11T}. Morphology is from NED and inclination $i$ from HyperLeda. The seeing is for the S7 observations \citep{2015ApJS..217...12D,2017ApJS..232...11T}}.    \label{tab:sample}
    \begin{tabular*}{\textwidth}{c @{\extracolsep{\fill}} cccccccc}
        \hline
        Galaxy & RA & $\delta$ & $z_{hel}$ & Type & $L(\oth)$ & Morph & $i$ & Seeing \\
        \hline
        NGC~1266    & 03:16:00.75 & $-$02:25:38.5 & 0.00724 & LINER & 40.9 & SB0 & 58$^\circ$& 1\farcs0 \\
        NGC~1808    & 05:07:42.34 & $-$37:30:47.0 & 0.00336 & SB & 39.9 & SABa & 83$^\circ$ & 1\farcs1 \\
        ESO~500-G34 & 10:24:31.45 & $-$23:33:10.7 & 0.01237 & SB$+$PSB$+$Sy2 & 41.1 & SB0/a & 75$^\circ$ & 1\farcs8 \\
        NGC~5728    & 14:42:23.89 & $-$17:15:11.0 & 0.00914 & Sy2 & 41.4 & SABa & 53$^\circ$ & 1\farcs2 \\
        ESO~339-G11 & 19:57:37.58 & $-$37:56:08.3 & 0.01907 & Sy2 & 42.2 & Sb & 74$^\circ$ & 1\farcs5 \\
        IC~5063     & 20:52:02.34 & $-$57:04:07.6 & 0.01129 & Sy2 & 41.8 & SA0 & 51$^\circ$ & 2\farcs0 \\
        IC~5169     & 22:10:09.98 & $-$36:05:19.0 & 0.01029 & SB$+$Sy2 & 40.4 & SAB0 & 84$^\circ$ & 1\farcs2 \\
        IC~1481     & 23:19:25.12 & $+$05:54:22.2 & 0.02036 & SB$+$Sy2 & 41.0 & S? & 29$^\circ$ & 1\farcs3 \\
        \hline
    \end{tabular*}
\end{table*}

In this work we use the S7 to probe spatially-resolved properties of
the cool, neutral gas in seven nearby AGN and one star-forming
galaxy. In Section \ref{sec:data} below, we describe the properties of
this subsample and our methods of data analysis. Two of these
galaxies, NGC~1266 and NGC~1808, were previously known to host
extensive dusty, neutral outflows based on \nad\ observations
\citep{1993AJ....105..486P,2012MNRAS.426.1574D}. As in other IFS
studies, these galaxies reveal blueshifted absorption lines. However,
NGC~1808 shows redshifted emission on the far side of the outflow
\citep{1993AJ....105..486P}. Deep, targeted IFS observations of a
handful of other nearby AGN have also detected spatially-resolved
\nad\ emission in outflowing gas
\citep{2015ApJ...801..126R,2019AA...623A.171P,2020MNRAS.494.5396B}.

We report below in Section~\ref{sec:data} that the wide FOV,
sensitivity, and spectral resolution of the S7 data combine to reveal
widespread \nad\ resonant line absorption and emission in our sample. In
Section~\ref{sec:results} we quantify the close connection between the
neutral gas and dust in these galaxies, and contextualize the results
within the deep dataset of Galactic absorbers and the much smaller
dataset of external galaxies with single-aperture \nad\ and
colour excess measurements. The current work expands significantly on our
knowledge of this connection within external galaxy disks.

In Section~\ref{sec:inflows_and_outflows}, we compare the kinematics
of these lines to those expected from simple disk rotation. We observe blue- or redshifted absorption and systemic or redshifted emission, as seen in previous studies, indicative of both inflows and outflows. For what appears to be the first time in this context, we also find numerous examples of inverse P-Cygni profiles: emission lines blueshifted from the
accompanying absorption. In Section~\ref{sec:discussion} we discuss the implications of these results, and summarize in Section~\ref{sec:summary}.

\section{DATA, ANALYSIS, AND MAPS} \label{sec:data}

We selected eight galaxies from the S7 survey
(\citealt{2007Ap&SS.310..255D,2010Ap&SS.327..245D};
Table~\ref{tab:sample}). S7 contains WiFeS data of 131 southern active
galaxies selected from
\citet{2006AA...455..773V,2010AA...518A..10V}. The galaxies in S7 lie at $z < 0.02$,
$\delta < 10^\circ$, and $|b| > 20^\circ$. For galaxies with 20~cm
radio continuum fluxes listed in the parent catalogue,
$f_\nu\ga20$~mJy is enforced \citep{2019AIPC.2109i0003S}. S7 galaxies primarily have nuclear optical spectral types of Seyfert, but there are some low-ionization nuclear emission-line region (LINER) galaxies and a handful classified as star-forming. They span the range of most SDSS AGN: $L(\oth) \sim10^{38}-10^{42}$~erg~s$^{-1}$, or
$L_\mathrm{bol} \sim10^{40}-10^{44}$~erg~s$^{-1}$
\citep{2006MNRAS.372..961K,2009AA...504...73L}. The galaxies in our
subsample were selected for strong, widespread \nad\ absorption and/or
emission by visual inspection of \nad\ linemaps created with QFitsView
\citep{2012ascl.soft10019O}. They are by no means a representative
subsample of S7, but were instead selected for their remarkable
resonant-line properties.

We reduced the data with PyWiFeS \citep{2014Ap&SS.349..617C,2014ascl.soft02034C} as part of the final S7 Data Release 2 (DR2), as described in \citet{2017ApJS..232...11T}. We apply Voronoi
binning \citep{2003MNRAS.342..345C,2012ascl.soft11006C} to the reduced data cubes in our subsample to enhance the signal-to-noise ratio (S/N) of the stellar 
continuum around \nad. The inputs to the Voronoi binning algorithm are the median flux and variance (per dispersion element) within a 30~\AA\ window centred on the redshifted 
\nad\ line. We set the target S/N per Voronoi bin to 8 or more and exclude spaxels with S/N$<$0.8.

The interstellar component of the \nad\ absorption and emission line
can be contaminated by stellar absorption. To accurately fit this
stellar component, the library of stellar spectra we use must be of
comparable or better spectral resolution than the data if the
intrinsic width of the observed stellar absorption lines is near the
spectral resolution. The S7 data have $R = 3000$ and 7000 in the blue
and red, respectively; these values are constant in wavelength over
the two output spectral ranges (350--570~nm and 540--700~nm). 
For the central regions of some of the most spatially-resolved galaxies in our data (e.g., NGC~1808), and in the outskirts of other galaxies, the stellar lines have observed widths comparable to the spectral resolution in the red part of the spectrum, where \nad\ arises.

The default spectral fits to S7 DR2 \citep{2017ApJS..232...11T} use the \citet{2005MNRAS.357..945G} single stellar population
synthesis models to fit the stellar continuum. These models have a
high dispersion (0.3~\AA/pixel) but their resolution is
unmeasured. The Indo-US library of empirical stellar spectra
\citep{2004ApJS..152..251V} has been shown to have a constant
resolution of 1.35~\AA\ \citep{2011AA...531A.109B}. It is beyond the
scope of this paper to measure the resolution of the
\citet{2005MNRAS.357..945G} library, but a one-to-one comparison with
the Indo-US library clearly indicates that the resolution of the
latter is higher. The Indo-US resolution is 3000--4000 in the blue
range of the S7 data, comparable to the data itself, but is only
4000--5000 in the red, vs. 7000 for the data.

\begin{figure*}
  \includegraphics[width=0.97\textwidth]{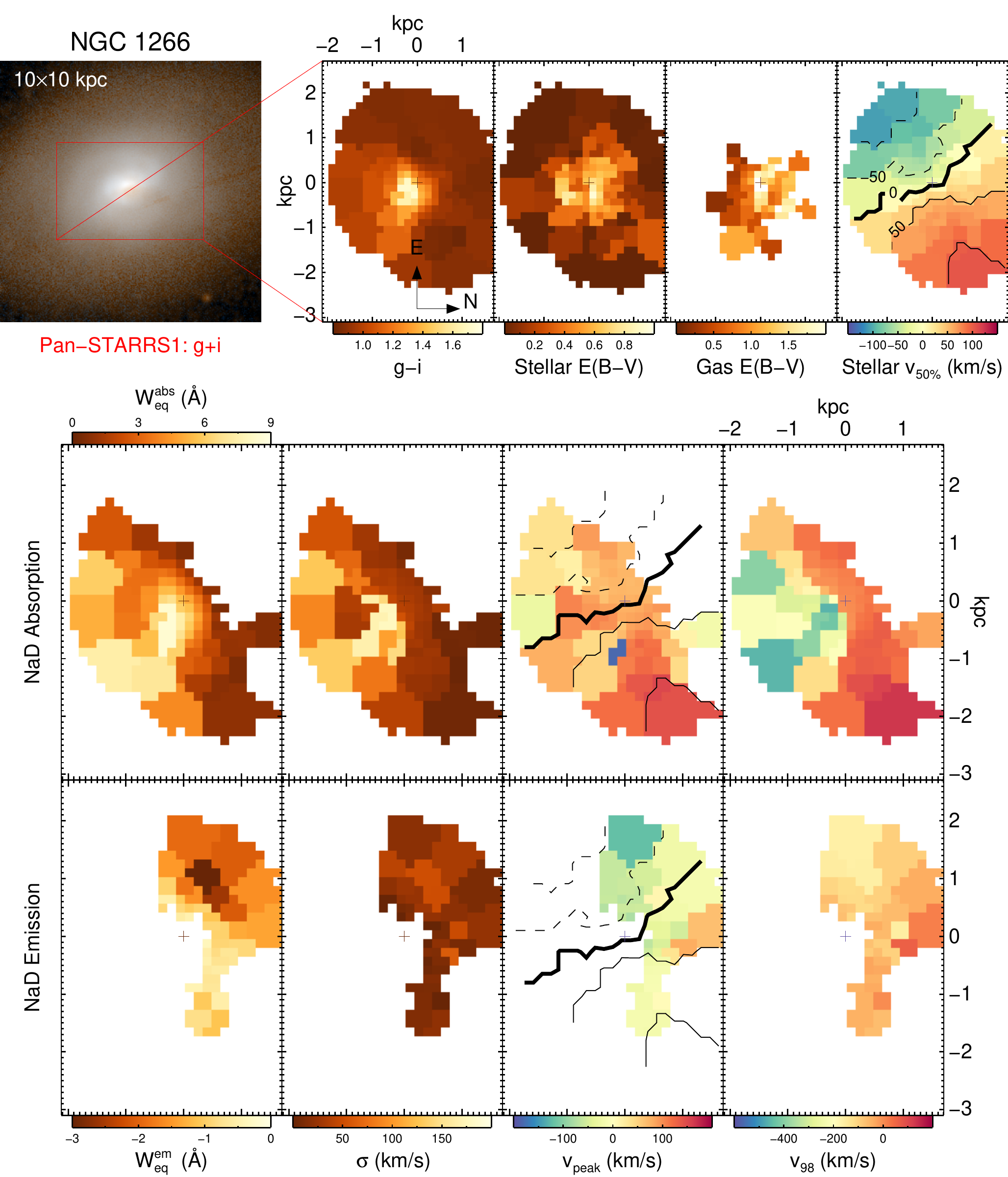}
  \caption{Images and maps of stellar and gas properties of each galaxy (Table~\ref{tab:imaging} and Section~\ref{sec:data}). \textit{Top row, left:} 2-colour image, with telescope and filters labeled below. The WiFeS FOV (25\arcsec$\times$38\arcsec) is outlined in red. \textit{Top row, middle and right:} Maps of broadband colour, stellar colour excess, ionized gas colour excess, and stellar velocity. Spaxel sizes are 1\farcs0. Broadband images are rebinned to match the Voronoi tessellation of the IFS data of each galaxy to compute colours. The stellar and gas properties are derived from fits to the data. The cross marks the continuum peak used to register the images and IFS maps, and distances from the peak are shown in kpc on the left and top axis of the colour map. Lines in the rightmost panel are isovelocity contours. \textit{Middle and bottom rows:} Equivalent width \weq, velocity dispersion $\sigma$, velocity at peak optical depth $v_\mathrm{peak}$, and maximum velocity \vtsig\ of \nad\ absorption and emission lines, in \AA\ and \kms. We derive these quantities from one-component fits except for the absorption lines in NGC~1266 and NGC~1808, for which we use two components in some spaxels. For one-component fits, $\sigma$ has the usual Gaussian definition and $\vtsig\equiv\vfifty-2\sigma$, where $\vfifty=v_\mathrm{peak}$ is the central velocity. For two-component fits, the velocities are derived from the cumulative velocity distribution in optical depth space, as described in the text. On the $v_\mathrm{peak}$ map, lines are the stellar isovelocity contours. \textit{The full set is available online.}}
  \label{fig:maps}
\end{figure*}

The Indo-US library contains 1273 spectra. However, we only need
representative stars across the range of stellar properties. We first
select spectra that have full coverage between 3465~\AA\ and 7000~\AA\
with any gaps in coverage smaller than 50~\AA, suitable for fitting
the S7 data. We also require measurements of effective temperature
$T_\mathrm{eff}$, gravity log $g$, and metallicity [Fe/H]. This
narrows the list to 1148 stars. Next, we evenly grid the resulting
spectra into bins of log $g$ and [Fe/H], with log $g$ ranging from 0
to 5 in bins of 0.25 and [Fe/H] ranging from $-$1 to 0.4 in units of
0.2. This captures the bulk of the distribution of these parameters in
the Indo-US catalog. We then add the tail of high-metallicity,
high-temperature (and high log $g$) stars by gridding in log
$T_\mathrm{eff}$ from 3.9 to 4.5 in units of 0.1 and over the same
range in metallicity. Finally, we take one star at random from each
bin and add it to our library. The final template library of 123 stars
samples a wide range in stellar properties.

The native pixel dispersions out of the S7 reduction pipeline are 0.76 and 0.44~\AA\ in the blue and red, respectively. We resample them to
0.88~\AA\ and stitch the two spectra at 5600~\AA. (The resampling and
stitching routine is part of the IFSRED library; \citealt{2014ascl.soft09004R}.)

\begin{figure*}
    \includegraphics[width=0.9\textwidth]{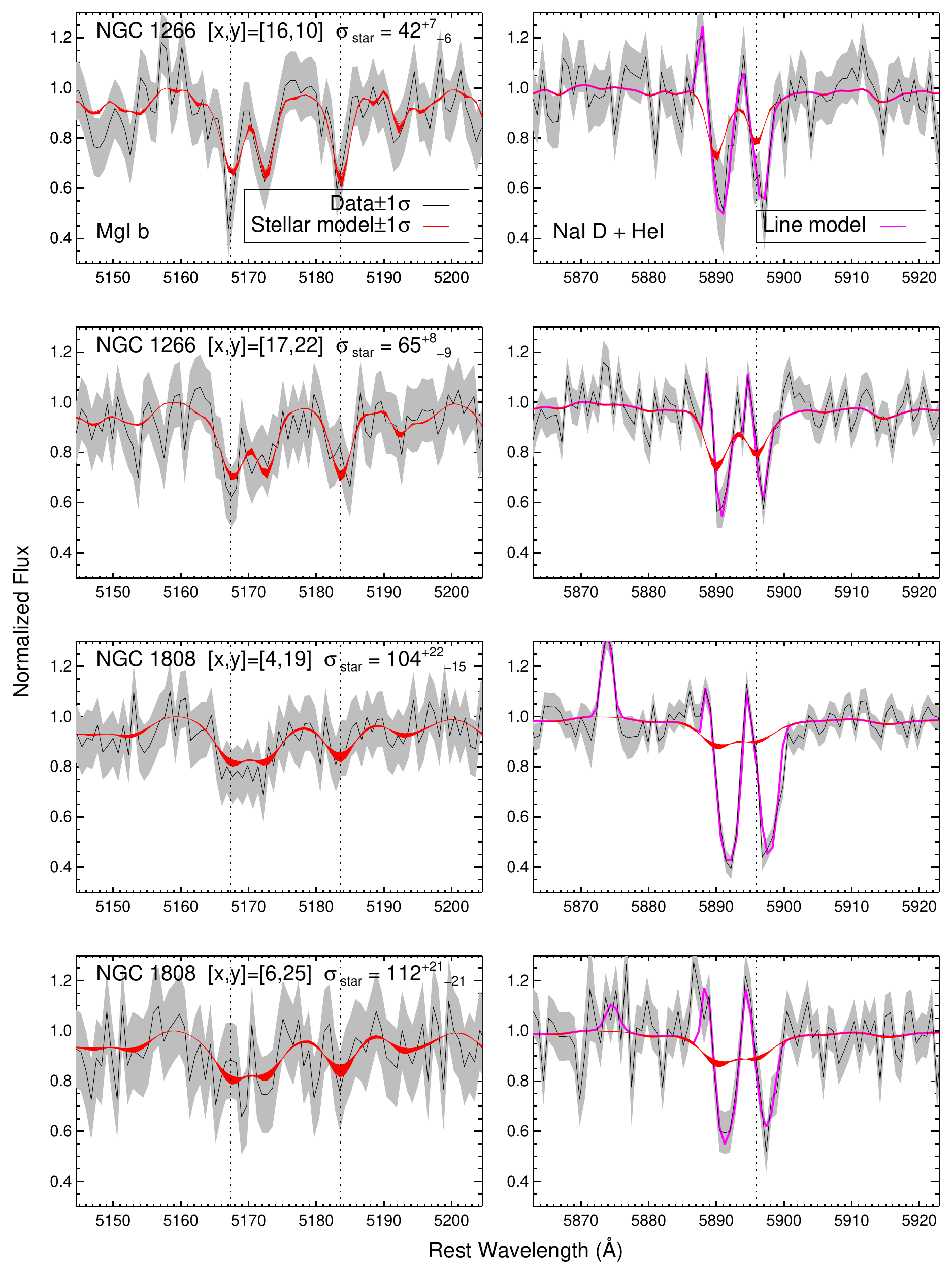}
    \caption{Four examples of inverse P Cygni profiles in our data. Each row represents a Voronoi bin in a data cube, with the bin labeled by one of its spaxels at column and row $[x,y]$, with spaxel $[1,1]$ at lower left and $[25,38]$ at upper right. \textit{Left}: Region around the \ion{Mg}{1}~b stellar lines; wavelengths are in the rest frame of the stars in the given spaxel. Black lines and grey shaded regions are the data and 1$\sigma$ error. Red area is the range of models given the best-fit stellar velocity dispersion $\sigma_\star$ and its 1$\sigma$ error. Dashed lines locate \ion{Mg}{1}~b. \textit{Right}:  Region around \nad. As at left, red area is the range of stellar models. Magenta lines show the best-fit \nad\ $+$ \ion{He}{1}~5876~\AA\ ISM model after subtraction of the stellar model. The expected locations of \nad\ and \ion{He}{1} based on the stellar velocity at each location are shown with vertical dashed lines. In each case the \nad\ emission doublet is blueshifted from the stellar velocity and the \nad\ absorption is redshifted.}
  \label{fig:inversePCygni}
\end{figure*}

We then fit the entire available spectral range of the Voronoi tessellated data cubes using IFSFIT \citep{2014ascl.soft09005R} and the Indo-US stellar templates. IFSFIT iteratively fits the stellar continuum and ionized gas emission lines. In each Voronoi bin, it does the following in order: masks potential interstellar features (emission lines and the \nad\ doublet); fits the stellar continuum with PPXF \citep{2004PASP..116..138C,2012ascl.soft10002C}; subtracts the continuum; and fits emission lines. It then repeats this process with more refined initial conditions based on the first fit.

Before performing the stellar continuum fits, we match the templates and data in spectral resolution to ensure that the fitted velocity dispersion is accurate. In the blue we convolve the template with a Gaussian for which the sigma equals the wavelength-dependent difference in resolution between the data and templates. In the red, we convolve the data to match the templates. Further fitting steps act on the unconvolved data after normalization by the continuum model. To account for residual calibration errors across the wide FOV, we include an additive 4th-order Legendre polynomial as part of the continuum model. For 5/8 galaxies, we increase the default polynomial order from 4 to 10 because of larger residuals. We assume a symmetric, Gaussian line-of-sight stellar velocity distribution ($MOMENT=2$ in the call to PPXF). We fit stellar color excess using the PPXF option to apply the \citet{2000ApJ...533..682C} reddening curve with $R_V = 4.05$ to the continuum. Finally, in each Voronoi bin, we fit errors in the recovered stellar parameters---velocity $v_\star$, velocity dispersion $\sigma_\star$ and colour excess \ebvs. To do so, we run a 100-iteration Monte Carlo simulation of the best-fit stellar continuum. Using the resulting distributions of best-fit $v_\star$, $\sigma_\star$ and \ebvs, we calculate the two-sided 1$\sigma$ errors on each parameter from the 34$\%$ interval on either side of the median.

The  stellar fits are of high quality. The median 1$\sigma$ errors in the stellar parameters, where the median is taken over the Voronoi bins in each data cube, range over 8--15~\kms\ in $v_\star$, 8--18~\kms\ in $\sigma_\star$ and 0.08--0.12 in \ebvs. The velocity errors are low compared to the spectral resolution of 100 (43)~\kms\ in the red (blue). The large wavelength range of the fit ensures that numerous spectral lines---both strong and weak---strongly constrain the fit and produce an excellent match between data and model throughout the dataset. This can be seen visually in the regular appearance of the $v_\star$ maps (Figure~\ref{fig:maps}) and in the good agreement of the data and model in the wavelength region around \ion{Mg}{1}~b (Figure~\ref{fig:inversePCygni}). The \nad\ and \ion{Mg}{1}~b stellar strengths are strongly correlated in most galaxies \citep[e.g.,][]{2013ApJS..208....7J}. Finally, visual comparison to published stellar velocity maps show a good match in NGC~1266 \citep{2011MNRAS.414.2923K}, NGC~1808 \citep{2017A&A...598A..55B}, and NGC~5728 \citep{2019ApJ...881..147S}.

We choose a redshift that produces the best antisymmetry around the  $v_\star=0$~\kms\ contour in the maps of $v_\star$ \citep{2011MNRAS.413..813C}. We then apply a heliocentric correction computed with the IDL routine BARYVEL \citep{1980A&AS...41....1S,1993ASPC...52..246L} These are reported in Table~\ref{tab:sample}, and have 1$\sigma$ errors of $\delta z = 0.00002$. In this study, all velocities for each galaxy are computed with respect to its systemic velocity.

Systematic errors in \nad\ stellar fitting could arise due to choice of stellar template. \citet{2010AJ....140..445C} find no significant effects on their study of \nad\ in SDSS between two choices of stellar models. We tested systematic uncertainties due to template choice by choosing a different subset of the Indo-US models; this has no effect. We also compared to fits using the \citet{2005MNRAS.357..945G} models; while less suited to this analysis due to the resolution mismatch (as we discuss above), they are used in the default S7 pipeline fits. Again, this change has almost no effect. The strength of the stellar \nad\ feature can also increase relative to \ion{Mg}{1}~b due to [Na/Fe] abundance variations. However, this effect is seen only in massive, early-type galaxies, not in the early-type spirals like those in our sample \citep{2013ApJS..208....7J}.

The deep S7 data contain a wealth of emission lines \citep{2017ApJS..232...11T}. These include the usual strong lines found in the rest-frame optical and fainter lines typically found in Seyfert nuclei. In particular, \ion{He}{1}~5876~\AA\ lies just blueward of \nad\ and can blend with the \nad\ profile. We discuss this line further below.

We constrain all visible lines in each bin to have the same velocity and velocity dispersion. Before fitting, we convolve each emission-line model with the wavelength-dependent spectral resolution of the data. We assume two velocity components as a baseline in most bins; components with S/N < 2$\sigma$ are dropped and
the spectrum is re-fit. A third component is added in a few bins
in NGC~5728 and IC~1481. In bins where both H$\alpha$ and H$\beta$ are detected, we use the Balmer decrement to calculate colour excess \ebvg\ assuming H$\alpha$/H$\beta=3.1$, appropriate for
Seyfert galaxies \citep{2006MNRAS.372..961K,2018ApJ...856...89T}, and
$R_V=4.05$.

Finally, we fit the \nad\ absorption and emission lines following procedures outlined in previous work \citep{2005ApJS..160...87R,2010ApJ...724.1430S,2015ApJ...801..126R}. We set the upper bound of the peak optical depth in the D$_1$ line to $\tau_{5896} = 5$; best-fit values of $\tau_{5896} = 5$ are considered lower limits. To avoid the inevitable degeneracy between emission and absorption line components in Voronoi bins where both arise, we first fit \nad\ emission and then absorption. This likely underestimates the equivalent width (\weq) and line width $\sigma$ in both emission and absorption where both are present and overlapping. Progress in modeling resonant line features with simulations or other gas phases \citep[e.g.,][]{2020MNRAS.494.5396B} may alleviate this difficulty. Each line model is convolved with the spectral resolution of the data before fitting. The lower limit for intrinsic linewidth $\sigma$ is set to 1~\kms, which is approximately the thermal value for Na atoms at 1000--6000~K. We fit one velocity component to every \nad\ emission line. The emission-line flux ratio $f(5890$~\AA$)/f(5896$~\AA) is allowed to vary between 1 and 2. We fit one component to every absorption line except in NGC~1266 and NGC~1808, where two absorption lines are fit in the nuclear regions and small off-nuclear regions. Fits of either absorption lines or emission lines with S/N $<2$ in equivalent width are discarded. To estimate errors, we perform 200-iteration Monte Carlo simulations in each Voronoi bin. As in computing stellar errors, we calculate the two-sided 1$\sigma$ errors on each parameter from the 34$\%$ interval on either side of the median.

We fit and remove \ion{He}{1}~5876~\AA\ emission during continuum and emission-line fitting for all galaxies but NGC~1266 and NGC~1808. Neither has a Seyfert optical spectral type (Table~\ref{tab:sample}) or low metallicity, so He emission lines are weak. NGC~1266 also has the broadest and deepest \nad\ absorption and NGC~1808 has the highest S/N data. In these two cases we fit  \ion{He}{1}~5876~\AA\ and \nad\ simultaneously. We constrain \ion{He}{1}~5876~\AA\ to have the same velocity and linewidth as other emission lines, and its flux is determined as part of the \nad\ fitting.

Example \nad\ fits are shown in Figure~\ref{fig:inversePCygni}. These Voronoi bins are chosen to showcase inverse P~Cygni profiles, which we discuss further in Sections~\ref{sec:inflows_and_outflows} and \ref{sec:discussion} below. They illustrate that the inverse P~Cygni profiles we detect are often visible prior to stellar continuum subtraction.

\begin{table}
    \centering
    \begin{threeparttable}
    \caption{Imaging data. The image sources are PS1 = Panoramic Survey Telescope and Rapid Response System Data Release 1 (Pan-STARRS1 DR1), CGS = Carnegie-Irvine Galaxy Survey \citep{2011ApJS..197...21H}, SSS = SkyMapper Southern Sky Survey DR3 \citep{2018PASA...35...10W}, and ATLAS = VLT Survey Telescope (VST) ATLAS Data Release 3 (DR3; \citealt{2015MNRAS.451.4238S}). Exposure times in $g$ and $i$ filters are given in s. Seeing for CGS, SSS, and ATLAS is measured by the survey; we estimate PS1 seeing from stars in the field. $i$ images are convolved to match $g$ seeing if they differ by more than 0\farcs1. The correction for Galactic reddening $\Delta^{g-i}_\mathrm{Gal}$ is given, from \citet{2011ApJ...737..103S} as implemented at \href{https://irsa.ipac.caltech.edu/applications/DUST/}{https://irsa.ipac.caltech.edu/applications/DUST/}. Other colour corrections $\Delta^{g-i}_\mathrm{other}$ are also listed. For SSS, the colour correction is from the SkyMapper to SDSS $g$ filters based on $g-i=1.0$. For ATLAS, the correction is from Vega to AB magnitudes \citep{2006MNRAS.367..454H}.}
    \label{tab:imaging}
    \begin{tabular*}{\columnwidth}{c @{\extracolsep{\fill}} ccccc}
        \hline
        Galaxy & Source & $t_\mathrm{exp}$ & Seeing & $\Delta^{g-i}_\mathrm{Gal}$ & $\Delta^{g-i}_\mathrm{other}$ \\
        \hline
        NGC~1266    & PS1   & 860/1800 & 1\farcs4 & -0.16 & --- \\
        NGC~1808\tnote{1} & CGS   & 360/180  & 1\farcs0 & -0.06 & --- \\
        ESO~500-G34 & PS1   & 654/1298 & 1\farcs2 & -0.09 & --- \\
        NGC~5728    & PS1   & 638/1118 & 1\farcs2 & -0.16 & --- \\
        ESO~339-G11 & SSS   & 5/100     & 3\farcs0 & -0.17 & 0.3 \\
        IC~5063     & SSS   & 100/100    & 2\farcs8 & -0.10 & 0.3 \\
        IC~5169     & ATLAS & 100/90   & 0\farcs8 & -0.03 & -0.46 \\
        IC~1481     & PS1   & 946/1920 & 1\farcs2 & -0.12 & --- \\
        \hline
    \end{tabular*}
    \begin{tablenotes}
        \item[1] Filters for NGC~1808 are $B$ and $I$ rather than $g$ and $i$; quantities in this table refer to $B$ and $I$ filters.
    \end{tablenotes}
    \end{threeparttable}
\end{table}

Broad-band optical images are available from various surveys
(Table~\ref{tab:imaging}). We use the deepest images that permit
calculation of $g-i$ colour as a proxy for stellar attenuation. (The
exception is NGC~1808, for which only $B-I$ colour is available.) We
sky-subtract and photometrically calibrate these images as
necessary, and convolve them to match the seeing of the WiFeS
observations if the WiFeS seeing is worse. As in \citet{2013ApJ...768...75R}, we register the images and data cube for each galaxy by computing the flux peak of the combined $g+i$ ($B+I$ for NGC~1808) image and an image created by stacking the data cube over all wavelengths. We compute the peaks using 2d, circular Moffat fits to the central pixels of each image with the IDL routine MPFIT2DPEAK \citep{2009ASPC..411..251M}. We then apply shifts to the images to align their peak to the data cube. The images, along with cutouts corresponding to the WiFeS FOV, are shown in Fig.~\ref{fig:maps}. This figure
also shows maps of colour, \ebvs,  \ebvg, and v$_\star$.

The distributions of \nad\ absorbing and emitting gas in each galaxy
are shown as \weq\ maps in the lower panels of
Fig.~\ref{fig:maps}. These panels also present the linewidth, peak
velocity, and 98\%\ velocity (\vtsig) of the cumulative velocity
distribution (CVDF) of absorbing and emitting gas. As in
\citet{2017ApJ...850...40R}, we construct the CVDF in optical depth
space for absorption lines and
compute velocities for which a specified percentage of the CVDF area
is redshifted from that velocity. For instance, 50\%\ of the CVDF lies
blue- and redward of \vfifty, and 98\%\ of the CVDF lies redward of
\vtsig. The velocity dispersion is defined as $\sigma \equiv (v_{16\%}-v_{84\%})/2$. 
For the \nad\ emission and single-component \nad\ absorption fits, the
central velocity \vfifty\ equals the velocity at the peak of the CVDF,
$v_\mathrm{peak}$ and $\vtsig\equiv\vfifty-2\sigma$.

\section{Spatially resolving the cool, neutral gas and dust
  connection} \label{sec:results}

The strong correlations between \nad\ absorption and colour excess derived from gas measurements
along stellar and QSO sightlines through the Milky Way's ISM and halo
have been studied for decades
\citep{1974ApJ...191..381H,1993AAS..100..107S,1994AA...289..539S,1994ApJ...436..152W,1994AJ....107.1022R,1997AA...318..269M,2000ApJ...544L.107W,2001ApJS..133..345W,2012MNRAS.426.1465P,2015MNRAS.452..511M}. \nad\
traces dusty, cool, neutral gas in the diffuse ISM through the disk;
this gas (traced by \ion{H}{1}, \ion{Na}{1}, and reddening) has scale
heights 0.4--0.5~kpc \citep{1994AA...289..539S}. These studies find
that \ion{Na}{1} absorption \weq\ and/or column density scale with
 \ebvg\ and/or N(\ion{H}{1}). There is significant scatter among
sightlines \citep{2000ApJ...544L.107W,2015MNRAS.452..511M} due to
varying ionization conditions and linewidths. High spectral resolution
finds complexes of low-$\sigma$ ($\sigma<10$~\kms) clouds
\citep{1974ApJ...191..381H,1993AAS..100..107S} that may comprise a
dusty and cold neutral medium of compact clouds
\citep{2019ApJ...887...89W,2019ApJ...886L..13P}. These Galactic
sightlines are thus primarily sensitive to low-to-moderate column
densities ($N($\ion{Na}{1}$)\la10^{12.5}$~cm$^{-2}$ and
$N($\ion{H}{1}$)\la10^{20.5}$~cm$^{-2}$) and colour excess values
$\ebvg \la 0.3$. Stacking analyses increase the ranges of $\ebv$
probed, but \weq\ still saturates at $\sim$1~\AA\ at  $\ebvg\ga0.2$
\citep{2012MNRAS.426.1465P}.

These correlations extend to external galaxies. Single-aperture
studies of both infrared- and optically-selected galaxies find strong
correlations between nuclear \ebvg\ and \nad\ absorption-line \weq\ \citep{1995ApJS...98..171V,2010AJ....140..445C}. Because these
sightlines trace larger projected areas, they encompass many more
clouds moving at a much wider range of velocities. They thus have
larger linewidths, equivalent widths, and column densities, but often
covering factors well below unity
\citep{2000ApJS..129..493H,2005ApJS..160...87R,2005ApJS..160..115R}.
A notable exception are the host galaxy absorbers in front of Type 1
QSOs \citep{2016ApJ...832....8B}, for which the background source has
a small projected area as seen by the host galaxy absorber (though not
perfectly point-like). These host galaxy absorbers also have
equivalent widths and column densities that correlate with reddening
\citep{2016ApJ...832....8B}.

Spatially-resolved correlations between \nad\ equivalent width and
continuum colour also exist in 2--3 nearby galaxies that are notably
dusty and have strong neutral outflows
\citep{2010ApJ...724.1430S,2013ApJ...768...75R,2015ApJ...801..126R}. Resonant
emission in \nad\ may in turn escape along low \ebv\ sightlines
\citep{2015ApJ...801..126R}. However, in at least one another system
(NGC 5626), there is no connection between \nad\ and dust attenuation
\citep{2017MNRAS.472.1286V}.

\begin{figure*}
  \includegraphics[width=\textwidth]{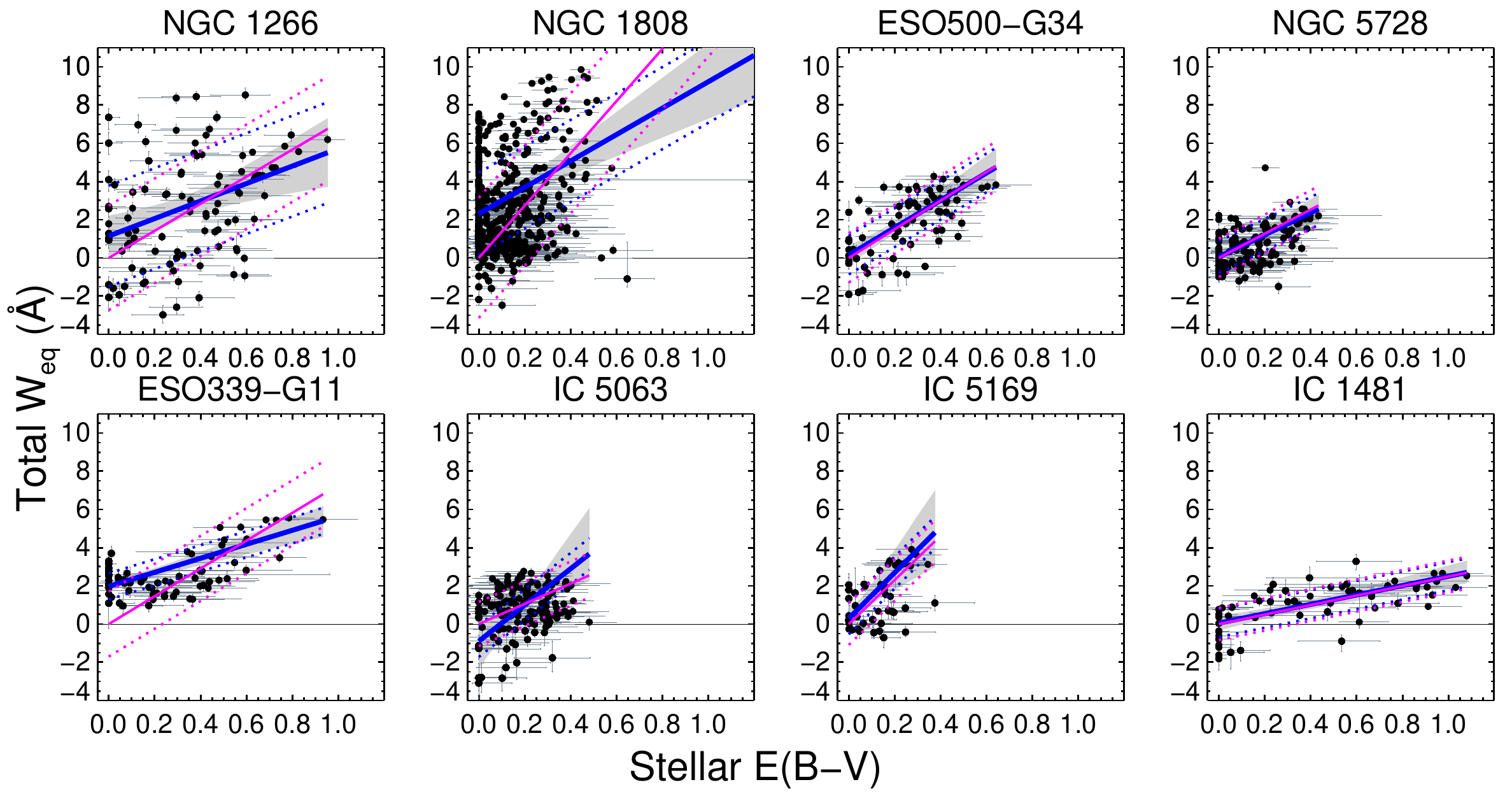}
  \caption{Total (absorption$+$emission) fitted \nad\ equivalent width
    vs. stellar colour excess, as measured from stellar continuum
    fitting. The solid blue and magenta lines are linear fits to the data
    with intrinsic Gaussian scatter using LINMIX\_ERR and MPFITEXY.} The gray area shows 2$\sigma$ deviations from the median LINMIX\_ERR fit after applying the posterior distributions of slopes and intercepts to the measured dependent
    variables. The dashed lines show the best fit lines plus fitted
    scatter. Regression models are shown only for those cases where the correlation is statistically significant at the 2$\sigma$ level. Data points are shown with 1$\sigma$ errors.
  \label{fig:weq_v_ebv_stel}
\end{figure*}

\begin{figure*}
  \includegraphics[width=\textwidth]{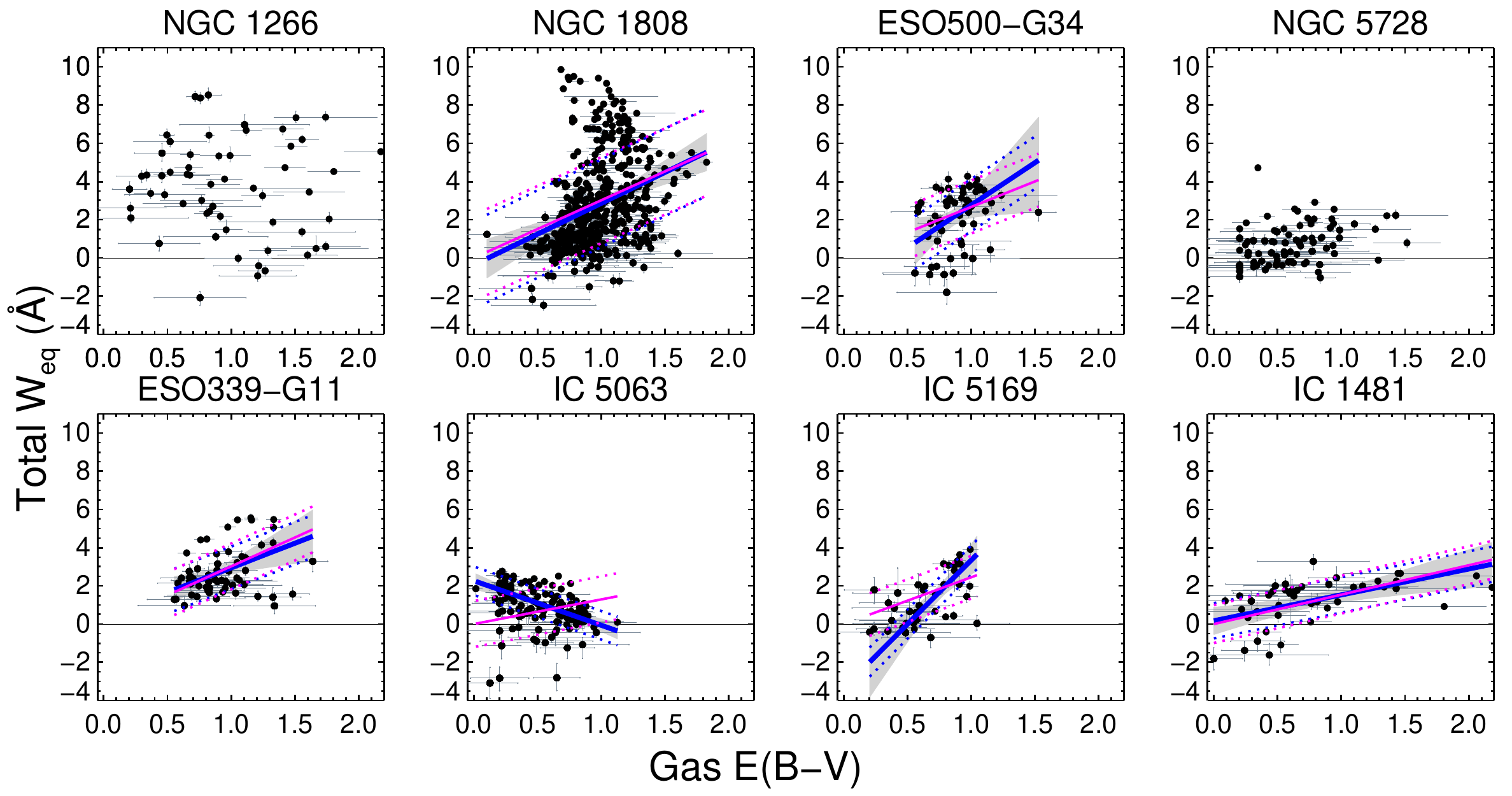}
  \caption{Total fitted \nad\ equivalent width vs. colour excess, as
    measured from the Balmer decrement. See Fig.
    \ref{fig:weq_v_ebv_stel} caption for more details.}
  \label{fig:weq_v_ebv_gas}
\end{figure*}

\begin{figure*}
  \includegraphics[width=\textwidth]{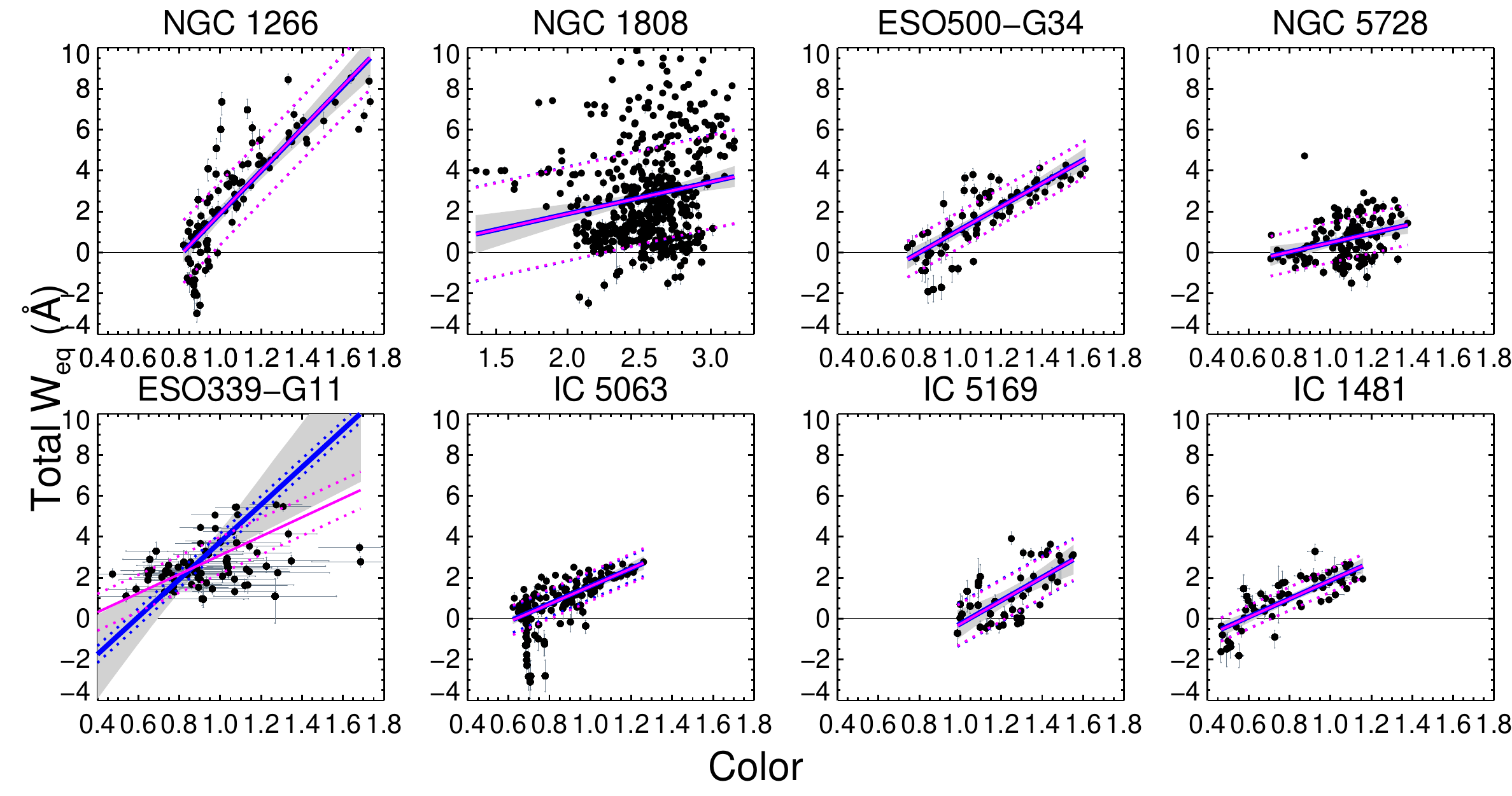}
  \caption{Total fitted \nad\ equivalent width vs. $g-i$ colour ($B-I$ colour for NGC~1808 only). See
    Fig. \ref{fig:weq_v_ebv_stel} caption for more details.}
  \label{fig:weq_v_col_gi}
\end{figure*}

\setlength{\extrarowheight}{5pt}
\begin{table}
    \centering
    \caption{Fits to \weq\ vs. stellar \ebv. Columns are sample size N; correlation coefficient $r_{xy}$; and regression intercept, slope, and intrinsic scatter. $p$-values are given in parentheses next to $r_{xy}$; many $p$-values are upper limits due to finite sampling of the correlation coefficient PDF. 1$\sigma$ errors are given for other quantities. The assumed model is $\weq = m[\ebv]+$\weq$_0+\delta\weq$, where m is the constant slope in \AA, \weq$_0$ is the equivalent width at $\ebv = 0$ in \AA, and $\delta$\weq\ (in \AA) is normally distributed. For each galaxy, the first row lists the results from a Bayesian regression with LINMIX\_ERR. The second row is from a regression with MPFITEXY with \weq$_0$ fixed to zero.}
    \label{tab:fits_weq_ebv_stel}
    \begin{tabular*}{\columnwidth}{l @{\extracolsep{\fill}} r  @{\extracolsep{\fill}}c  @{\extracolsep{\fill}}c @{\extracolsep{\fill}}c @{\extracolsep{\fill}}c}
        \hline
        Galaxy & N & $r_{xy}$ & \weq$_0$ & $m$ & $\delta$\weq\\[0.1cm]
        \hline
  NGC 1266  &  99  &  0.37 (0.0001)  &  1.14$_{-  0.51}^{+  0.51}$   &  4.58$_{-  1.36}^{+  1.35}$   &  2.64$_{-  0.19}^{+  0.23}$  \\
            &      &        &0.0  &  7.08$\pm$0.65  &  2.73$\pm$0.19 \\
  NGC 1808  & 503  &  0.40 (<6e-5)  &  2.31$_{-  0.11}^{+  0.11}$   &  6.92$_{-  1.02}^{+  1.00}$   &  2.15$_{-  0.07}^{+  0.08}$  \\
            &      &        &0.0  & 13.68$\pm$2.84  &  3.14$\pm$0.09 \\
ESO500-G34  &  73  &  0.77 (<6e-5)  &  0.15$_{-  0.24}^{+  0.23}$   &  7.17$_{-  0.85}^{+  0.94}$   &  1.01$_{-  0.10}^{+  0.14}$  \\
            &      &        &0.0  &  7.50$\pm$0.49  &  1.29$\pm$0.11 \\
  NGC 5728  & 130  &  0.59 (<6e-5)  &  0.05$_{-  0.13}^{+  0.15}$   &  5.77$_{-  1.16}^{+  0.92}$   &  0.83$_{-  0.07}^{+  0.08}$  \\
            &      &        &0.0  &  6.33$\pm$0.78  &  1.00$\pm$0.10 \\
ESO339-G11  &  74  &  0.79 (<6e-5)  &  1.98$_{-  0.12}^{+  0.12}$   &  3.67$_{-  0.49}^{+  0.46}$   &  0.68$_{-  0.09}^{+  0.12}$  \\
            &      &        &0.0  &  7.28$\pm$0.33  &  1.70$\pm$0.10 \\
   IC 5063  & 148  &  0.62 (<6e-5)  & -0.90$_{-  0.64}^{+  0.47}$   &  9.49$_{-  2.52}^{+  3.44}$   &  0.83$_{-  0.09}^{+  0.10}$  \\
            &      &        &0.0  &  5.29$\pm$0.53  &  1.16$\pm$0.06 \\
   IC 5169  &  47  &  0.78 (<6e-5)  &  0.30$_{-  0.22}^{+  0.24}$   & 11.99$_{-  2.65}^{+  2.93}$   &  0.83$_{-  0.16}^{+  0.24}$  \\
            &      &        &0.0  & 11.55$\pm$1.37  &  1.09$\pm$0.16 \\
   IC 1481  &  59  &  0.74 (<6e-5)  &  0.05$_{-  0.17}^{+  0.17}$   &  2.45$_{-  0.36}^{+  0.37}$   &  0.73$_{-  0.09}^{+  0.12}$  \\
            &      &        &0.0  &  2.49$\pm$0.20  &  0.84$\pm$0.07 \\
        \hline
    \end{tabular*}
\end{table}

\begin{table}
    \centering
    \caption{Fits to \weq\ vs. gas \ebv. See Table~\ref{tab:fits_weq_ebv_stel} for more details.}
    \label{tab:fits_weq_ebv_gas}
    \begin{tabular*}{\columnwidth}{l @{\extracolsep{\fill}} r  @{\extracolsep{\fill}}c  @{\extracolsep{\fill}}c @{\extracolsep{\fill}}c @{\extracolsep{\fill}}c}
        \hline
        Galaxy & N & $r_{xy}$ & \weq$_0$ & $m$ & $\delta$\weq\\[0.1cm]
        \hline
   NGC 1266  &  59  & -0.02 (0.5)  &  3.80$_{-  0.88}^{+  0.91}$   & -0.09$_{-  0.84}^{+  0.85}$   &  2.60$_{-  0.22}^{+  0.29}$  \\
            &      &   &   0.0  &  3.09$\pm$0.37  &  3.07$\pm$0.21 \\
  NGC 1808  & 458  &  0.29 (<6e-5)  & -0.37$_{-  0.56}^{+  0.57}$   &  3.23$_{-  0.56}^{+  0.55}$   &  2.29$_{-  0.08}^{+  0.08}$  \\
            &      &   &   0.0  &  3.02$\pm$0.11  &  2.25$\pm$0.09 \\
ESO500-G34  &  59  &  0.45 (0.01)  & -1.68$_{-  1.54}^{+  1.67}$   &  4.45$_{-  1.90}^{+  1.71}$   &  1.37$_{-  0.15}^{+  0.19}$  \\
            &      &   &   0.0  &  2.67$\pm$0.20  &  1.39$\pm$0.13 \\
  NGC 5728  &  82  &  0.24 (0.05)  &  0.04$_{-  0.40}^{+  0.42}$   &  1.14$_{-  0.65}^{+  0.64}$   &  1.03$_{-  0.08}^{+  0.10}$  \\
            &      &   &   0.0  &  1.30$\pm$0.17  &  1.02$\pm$0.12 \\
ESO339-G11  &  59  &  0.46 (0.002)  &  0.36$_{-  0.78}^{+  0.77}$   &  2.58$_{-  0.85}^{+  0.89}$   &  1.11$_{-  0.11}^{+  0.13}$  \\
            &      &   &   0.0  &  3.02$\pm$0.21  &  1.20$\pm$0.11 \\
   IC 5063  & 140  & -0.58 (<6e-5)  &  2.27$_{-  0.22}^{+  0.21}$   & -2.33$_{-  0.34}^{+  0.35}$   &  0.75$_{-  0.06}^{+  0.07}$  \\
            &      &   &   0.0  &  1.29$\pm$0.15  &  1.20$\pm$0.06 \\
   IC 5169  &  43  &  0.84 (<6e-5)  & -3.35$_{-  1.17}^{+  0.94}$   &  6.69$_{-  1.26}^{+  1.51}$   &  0.76$_{-  0.14}^{+  0.18}$  \\
            &      &   &   0.0  &  2.45$\pm$0.27  &  1.11$\pm$0.08 \\
   IC 1481  &  43  &  0.59 (0.0008)  &  0.17$_{-  0.34}^{+  0.34}$   &  1.35$_{-  0.38}^{+  0.37}$   &  0.92$_{-  0.12}^{+  0.16}$  \\
            &      &   & 0.0  &  1.54$\pm$0.16  &  0.99$\pm$0.09 \\
        \hline
    \end{tabular*}
\end{table}

\begin{table}
    \centering
    \begin{threeparttable}
    \caption{Fits to \weq\ vs. $g-i$ colour. See Table~\ref{tab:fits_weq_ebv_stel} for more details. In this case, $(g-i)_0$ is the x-intercept computed from the fitted
    y-intercept, and the y-intercept is not fixed to zero in either regressions.}
    \label{tab:fits_weq_col}
    \begin{tabular*}{\columnwidth}{l @{\extracolsep{\fill}} r  @{\extracolsep{\fill}}c  @{\extracolsep{\fill}}c @{\extracolsep{\fill}}c @{\extracolsep{\fill}}c}
        \hline
        Galaxy & N & $r_{xy}$ & $(g-i)_0$ & $m$ & $\delta$\weq\\[0.1cm]
        \hline
  NGC 1266  &  99  &  0.83 (<6e-5)  &  0.82$_{-  0.09}^{+  0.09}$   & 10.36$_{-  0.69}^{+  0.72}$   &  1.53$_{-  0.11}^{+  0.13}$  \\
            &      &        &  0.82$\pm$0.10  & 10.40$\pm$0.80  &  1.54$\pm$0.14 \\
  NGC 1808  & 503  &  0.19 (<6e-5)  &  0.78$_{-  0.63}^{+  0.65}$   &  1.56$_{-  0.37}^{+  0.37}$   &  2.30$_{-  0.07}^{+  0.08}$  \\
            &      &        &  0.78$\pm$0.72  &  1.55$\pm$0.42  &  2.27$\pm$0.08 \\
ESO500-G34  &  73  &  0.83 (<6e-5)  &  0.80$_{-  0.12}^{+  0.12}$   &  5.64$_{-  0.48}^{+  0.49}$   &  0.89$_{-  0.08}^{+  0.10}$  \\
            &      &        &  0.80$\pm$0.11  &  5.63$\pm$0.40  &  0.87$\pm$0.10 \\
  NGC 5728  & 130  &  0.32 (<6e-5)  &  0.79$_{-  0.36}^{+  0.34}$   &  2.27$_{-  0.58}^{+  0.61}$   &  0.98$_{-  0.06}^{+  0.07}$  \\
            &      &        &  0.79$\pm$0.35  &  2.28$\pm$0.60  &  0.97$\pm$0.09 \\
ESO339-G11  &  74  &  0.93 (<6e-5)  &  0.59$_{-  0.23}^{+  0.23}$   &  9.15$_{-  1.96}^{+  1.92}$   &  0.41$_{-  0.13}^{+  0.22}$  \\
            &      &        &  0.33$\pm$0.56  &  4.64$\pm$2.77  &  0.90$\pm$0.24 \\
   IC 5063  & 148  &  0.74 (<6e-5)  &  0.63$_{-  0.09}^{+  0.09}$   &  4.32$_{-  0.34}^{+  0.34}$   &  0.64$_{-  0.05}^{+  0.06}$  \\
            &      &        &  0.64$\pm$0.09  &  4.36$\pm$0.32  &  0.72$\pm$0.08 \\
   IC 5169  &  47  &  0.68 (<6e-5)  &  1.04$_{-  0.30}^{+  0.31}$   &  5.67$_{-  1.06}^{+  1.04}$   &  1.04$_{-  0.11}^{+  0.15}$  \\
            &      &        &  1.05$\pm$0.25  &  5.68$\pm$0.85  &  0.99$\pm$0.10 \\
   IC 1481  &  59  &  0.85 (<6e-5)  &  0.58$_{-  0.10}^{+  0.10}$   &  4.45$_{-  0.42}^{+  0.44}$   &  0.57$_{-  0.07}^{+  0.09}$  \\
            &      &        &  0.59$\pm$0.09  &  4.49$\pm$0.40  &  0.57$\pm$0.06 \\
        \hline
    \end{tabular*}
    \begin{tablenotes}
        \item[1] Filters for NGC~1808 are $B$ and $I$ rather than $g$ and $i$; quantities in this table refer to $B$ and $I$ filters.
    \end{tablenotes}
    \end{threeparttable}
\end{table}

For one well-resolved source, \citet{2015ApJ...801..126R} use a spatial model of dust and \nad\ that separates the emission and absorption line contributions to the equivalent width in each spaxel. This sample is of larger size and our interest lies in looking for correlations with model-independent observables. Here we choose a simpler approach and use total equivalent width ($\weqa+\weqe$) as a proxy that includes both within each Voronoi bin. However, the number of points in which both are present is a small fraction of the total and the relationships are driven by points in which one or the other is present.

Our data allow us to quantify the spatially-resolved
connection between \nad\ equivalent width and dust across the disks in our subsample. Fig.~\ref{fig:weq_v_ebv_stel}--\ref{fig:weq_v_col_gi} display the relationship between total equivalent width ($\weqa+\weqe$) and stellar \ebv, gas \ebv, and $g-i$ colour for each galaxy. There is significant intrinsic scatter in these plots unaccounted for by the measurement errors, so as in \citet{2017ApJ...850...40R} we compute Bayesian linear regressions using LINMIX\_ERR \citep{2007ApJ...665.1489K}. Based on samples of the posterior probability distribution functions (PDFs), we list the median and $p$-value for the correlation coefficient and median and 1$\sigma$ errors for intercept, slope and instrinsic scatter in Tables~\ref{tab:fits_weq_ebv_stel}--\ref{tab:fits_weq_col}. Many $p$-values are upper limits due to finite sampling of the correlation coefficient PDF.

For comparison, we also compute regressions using MPFITEXY \citep{2010MNRAS.409.1330W} and find the resulting errors in slope and intrinsic scatter using a simple 1000-sample bootstrap \citep{2010arXiv1008.4686H}. We enable the option to adjust the intrinsic scatter to produce $\chi_r \sim 1$ \citep{2006MNRAS.373.1125B}. Here, we also fix the y-intercept to zero for \weq\ vs. \ebv\ for comparison to the unfixed values in the LINMIX\_ERR fits. It is unclear whether or not total \weq\ should be zero at $\ebv = 0$; \citet{2010AJ....140..445C} find that $\weq < 0$ at $\ebv = 0$, while in \citet{2012MNRAS.426.1465P} the intercept is consistent with zero. For the \weq\ vs. $g-i$ fit, we list the x-intercept, which should relate to the intrinsic colour of the galaxy at low colour excess.

The neutral gas and dust show statistically significant correlations throughout; $r_{xy}>0$ at the 95.5\%\ (99.99\%) level in 22/24 (18/24) fits. The significance and strengths of these correlations are highest for \weq\ vs. \ebvs\ and $g-i$. We find a median $r_{xy}$ of 0.74 and 0.83 for \weq\ with \ebvs\ and $g-i$, respectively, vs. 0.45 for \weq\ with \ebvg. There is intrinsic scatter in these correlations, however, with a median value across the sample (and across fitting methods) of 1.0~\AA.

The results from the two fitting methods produce very consistent results, in particular for the $g-i$ fits. Differences appear largely from fixing the y-intercept in MPFITEXY in the \ebv\ fits, though in cases where the fitted intercept with LINMIX\_ERR is near zero the slopes are consistent. The covariance between slope and intercept is evident in that, for any galaxy, larger best-fit slopes result from the fit (LINMIX\_ERR or MPFITEXY) with smaller intercept, and vice versa. The parameter errors are typically larger from LINMIX\_ERR than those from the bootstrap errors with MPFITEXY, while the fitted intrinsic scatter is in turn lower. This points to differences in how the methods (LINMIX\_ERR and MPFITEXY with bootstrap errors) treat the tradeoff between parameter errors and intrinsic scatter.

When the median and standard deviation in the slope is taken over all galaxies for each type of correlation and each fitting method, the methods differ primarily in that the MPFITEXY method produces much lower variance for \weq\ vs. \ebvg. Taking the method with the lowest variance for each type of correlation, the values for median slope and standard deviation across the 8-galaxy sample are (6.9$\pm$3.1)~\AA, (2.7$\pm$0.8)~\AA, and (4.6$\pm$2.7)~\AA\ for \weq\ vs. \ebvs, \ebvg, and $g-i$, respectively. As a physical sanity check, the ratio of the median slopes from the fits to \ebvg\ and \ebvs, $m(\ebvg)/m(\ebvs$, is 0.39$\pm$0.07. This ratio of slopes is consistent with the average ratio of stellar to gas colour excess, $\ebvs/\ebvg$, for which the canonical value is 0.44$\pm$0.03 \citep{2000ApJ...533..682C,2013ApJ...771...62K}, though this can vary from galaxy to galaxy and within galaxies \citep{2020MNRAS.495.2305G}.

For half of the sample (ESO~500-G34, ESO 339-G11, IC~5169, and IC~1481) the correlations between \weq\ and dust measure (\ebvs, \ebvg, or $g-i$) are consistently strong ($r_{xy} \ga 0.5$) and significant among all three dust measures. In other galaxies the situation is less clear. In NGC~1266, the strongest and most signficant correlation, with the lowest scatter in regression, is \weq\ vs. $g-i$. The \weq\ vs. \ebvs\ correlation is weaker and the fit has larger scatter, while \weq\ and \ebvg\ show no correlation at all. NGC~1266 also has the steepest fitted slope of any galaxy in \weq\ vs. colour space. In IC~5063, \weq\ correlates strongly and significantly with \ebvs\ and colour, but has a strong anti-correlation with \ebvg. The colour map reflects the strong dust lane running NW to SE across the near side (NE) of the galaxy, which is also where the \nad\ \weq\ is highest, while \ebvg\ shows strong obscuration away from the dust lane and low obscuration in the dust lane (Figure~\ref{fig:maps}$f$). NGC~1808 has weak (though significant) correlations in all three relationships but very large scatter ($r = 0.2-0.4$ and $\delta\weq > 2.3$~\AA). It is the nearest and best-resolved galaxy in our sample, and the high S/N data record spatially-resolved patterns that are not best described with linear relationships. Finally, in NGC~5728 there is a region of scattered AGN light \citep{1996ApJ...466..169C} that skews the colour map, so that the stellar \ebv\ correlation is strongest.

As a final sanity check, the x-intercepts of the $g-i$ fits should be consistent with observed intrinsic stellar colours. The S7 galaxies have masses $M_\star = 10^{10-11}~M_\odot$ and lie mostly in the blue sequence or green valley. Integrated unattenuated colours from the Galaxy And Mass Assembly (GAMA) survey are in the range 0.2--0.8 for $M_\star = 10^{10-11}~M_\odot$ blue sequence galaxies, and a bit higher for the green valley \citep{2015MNRAS.446.2144T}. Our measured $(g-i)_0$ are consistent with, but fall toward the high end of, these values, with a median of 0.8 and a range of 0.3--1.0.

\begin{figure}
  \includegraphics[width=\columnwidth]{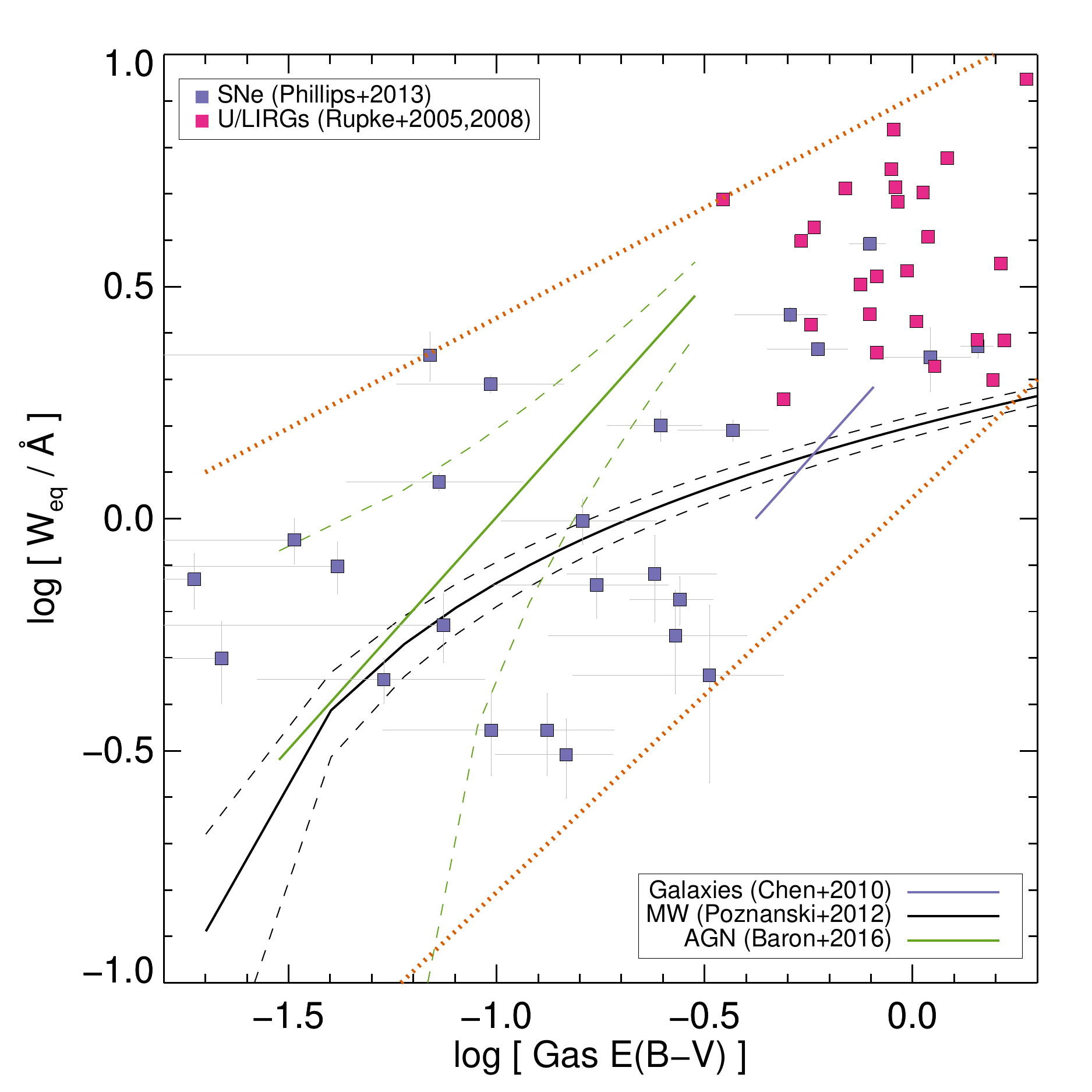}
  \caption{\nad\ absorption equivalent width vs. colour excess of ionized gas in the
    Milky Way and external galaxies. Purple and magnenta squares show
    individual sightlines to supernovae \citep{2013ApJ...779...38P}
    and infrared-luminous galaxies
    \citep{2005ApJS..160...87R,2008ApJ...674..172R}. Lines show fits
    to the relationship between these quantities for SDSS star-forming
    galaxies \citep[purple solid line]{2010AJ....140..445C}, AGN
    \citep[green solid line]{2016ApJ...832....8B}, and Milky Way
    sightlines to background galaxies and quasars \citep[black solid
    line]{2012MNRAS.426.1465P}. Dashed lines delineate 1$\sigma$
    uncertainties in the fits, while gray lines show 1$\sigma$ errors
    in data points. The orange dotted lines show the approximate range
    of the data and fits.}
  \label{fig:weq_abs_v_ebv_gas_other}
\end{figure}

\begin{figure}
  \includegraphics[width=\columnwidth]{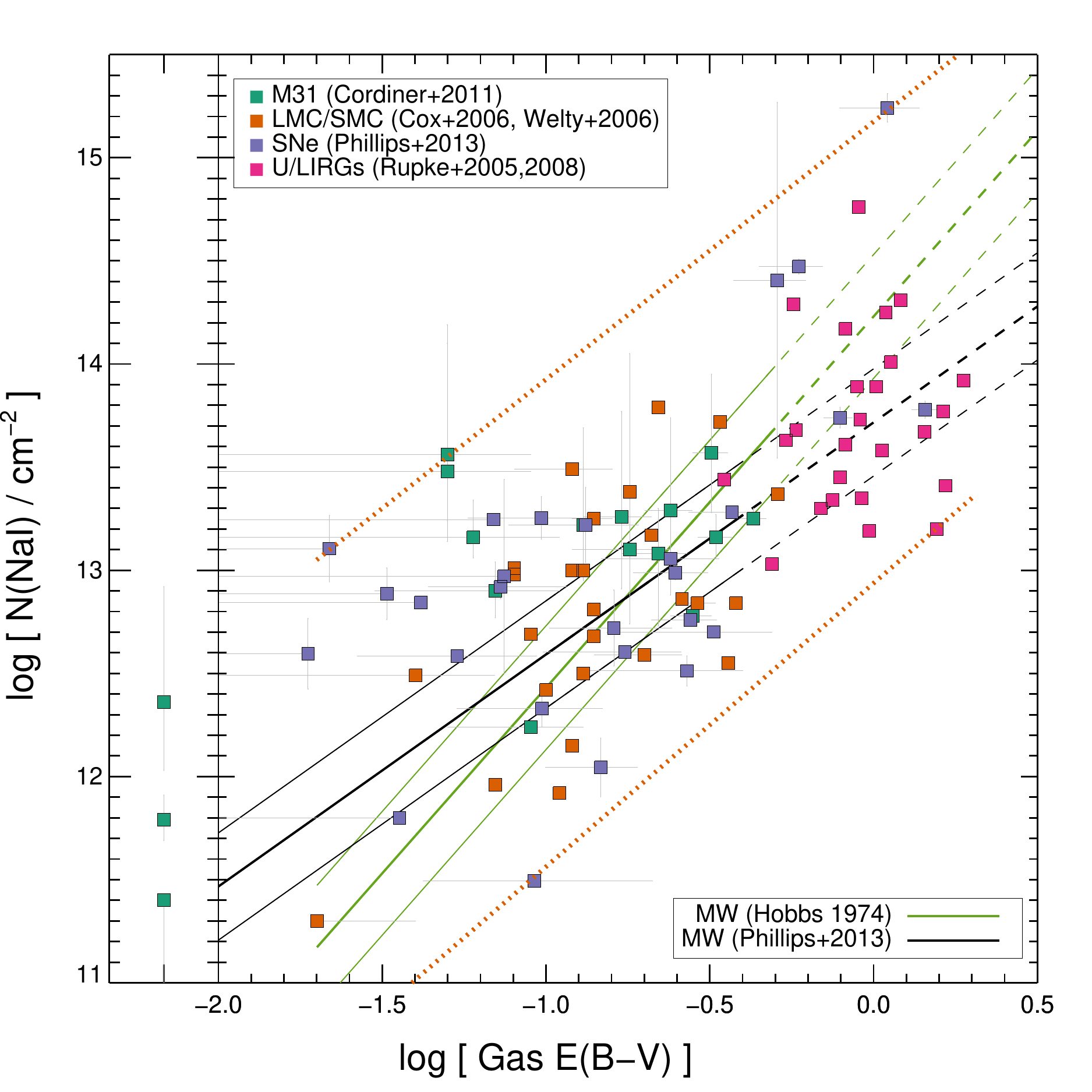}
  \caption{\nad\ absorption column density vs. colour excess of ionized gas in the
    Milky Way and external galaxies. In addition to the sightlines
    through infrared-luminous galaxies (pink squares) and galaxies
    containing supernovae (purple) shown in Fig.
    \ref{fig:weq_abs_v_ebv_gas_other}, sightlines to bright stars in
    M31 (green) and the LMC/SMC (orange) are shown. Green and black
    lines show fits to Milky Way data $\pm$1$\sigma$
    \citep{1974ApJ...191..381H,2013ApJ...779...38P}, and dashed lines
    are extensions of these relationships to higher \ebvg. Dotted orange lines show the approximate range of the
    data.}
  \label{fig:nnai_abs_v_ebv_gas_other}
\end{figure}

To contextualize these results, we explore where our data lie in
\weqa\ and $N$(\ion{Na}{1}) vs. \ebv$_\mathrm{gas}$ space. We calculate column density as in \citet{2005ApJS..160...87R}, and where the best-fit D$_1$ optical depth is a lower limit ($\tau_{5896} \ge 5$) we consider the column density to be a lower limit. We
compare to the extensive measurements in the Milky Way, which largely
probe low extinctions, and the limited set of sightlines in external
galaxies that reach higher \ebv. The most recent study of MW
sightlines uses both stars and background galaxies/quasars
\citep{2012MNRAS.426.1465P} to derive a high-precision average \weqa\
vs. \ebv\ (Fig.~\ref{fig:weq_abs_v_ebv_gas_other}). External galaxy
sightlines (backlit by point sources like AGN/supernovae or nuclear
starbursts) straddle this relationship but often lie above it by
factors up to 0.6~dex, particlarly at $\ebv \ga 0.5$
\citep{2005ApJS..160...87R,2008ApJ...674..172R,2010AJ....140..445C,2013ApJ...779...38P,2016ApJ...832....8B}. Two fits to MW stellar measurements of $N$(\ion{Na}{1}) and \ebv, despite being made 40 years apart, yield roughly the same result (Fig.~\ref{fig:nnai_abs_v_ebv_gas_other}). Most external galaxy measurements lie roughly in the extrapolation of these fits, with a possible tendency to scatter above rather than below them.

\begin{figure*}
  \includegraphics[width=\textwidth]{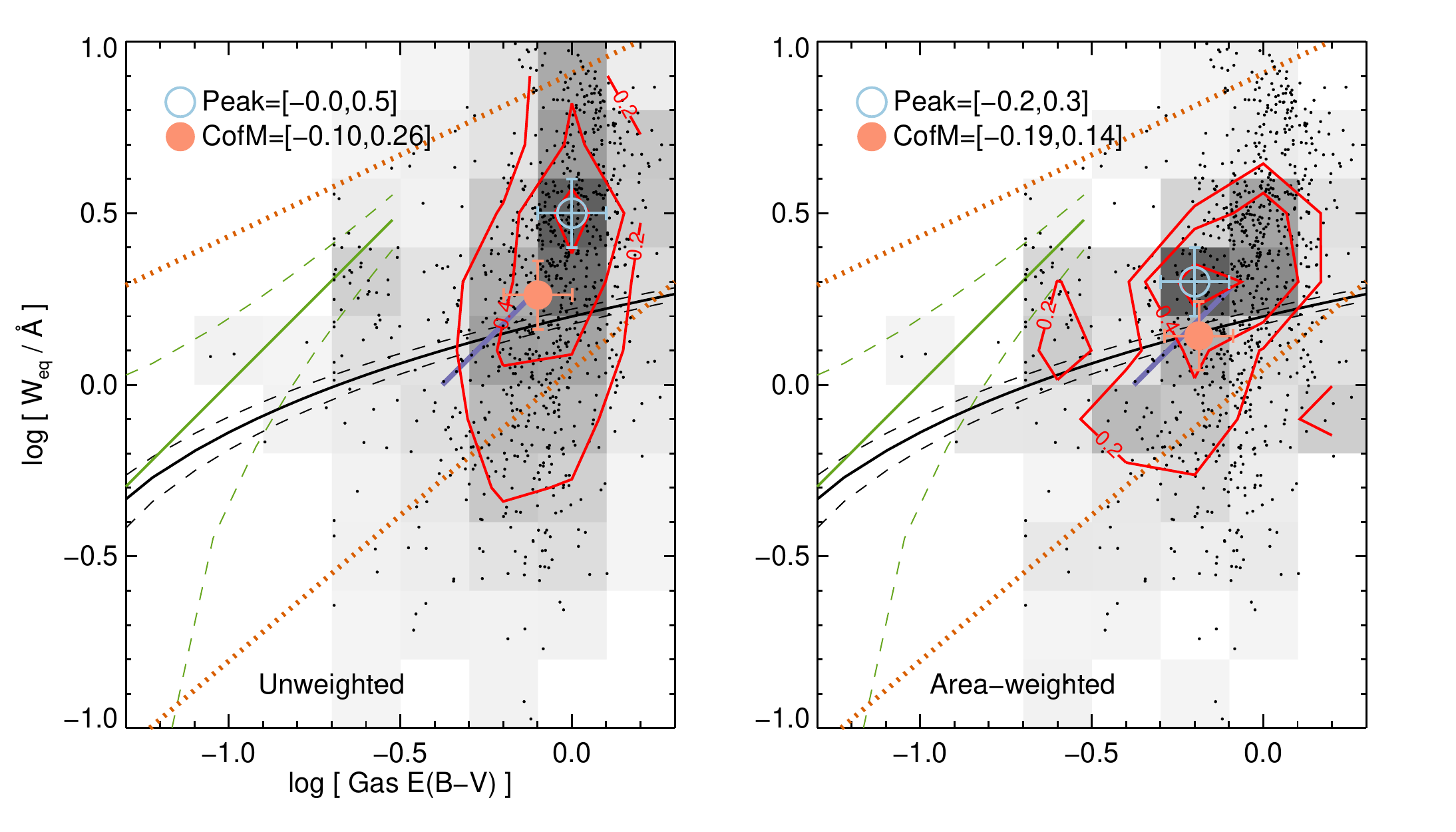}
  \caption{\nad\ absorption equivalent width vs. colour excess from gas measurements in
    our sample, compared to MW and other external galaxy
    sightlines. Black circles are individual Voronoi bins. The spatial bins are binned in \weq\ vs. \ebv\ space and the grey shading of each bin represents the number of Voronoi bins within a plot bin. The binning on the left is unweighted, while on the right bins are first weighted by the projected area they subtend on the galaxy. Red contour lines of 0.2, 0.4, and 0.8 represent (weighted) number of spaxels in each Voronoi bin divided by the value in the peak bin. The light blue open and light red filled circles are the peak and "center-of-mass" of the binned data. The other lines are as described in Fig. \ref{fig:weq_abs_v_ebv_gas_other}. Our data are on average consistent with single-apertures sightlines through the Milky Way and external galaxies, though they probe to both lower and higher \weq\ values for a given \ebv.}
  \label{fig:weq_abs_v_ebv_gas}
\end{figure*}

\begin{figure*}
  \includegraphics[width=\textwidth]{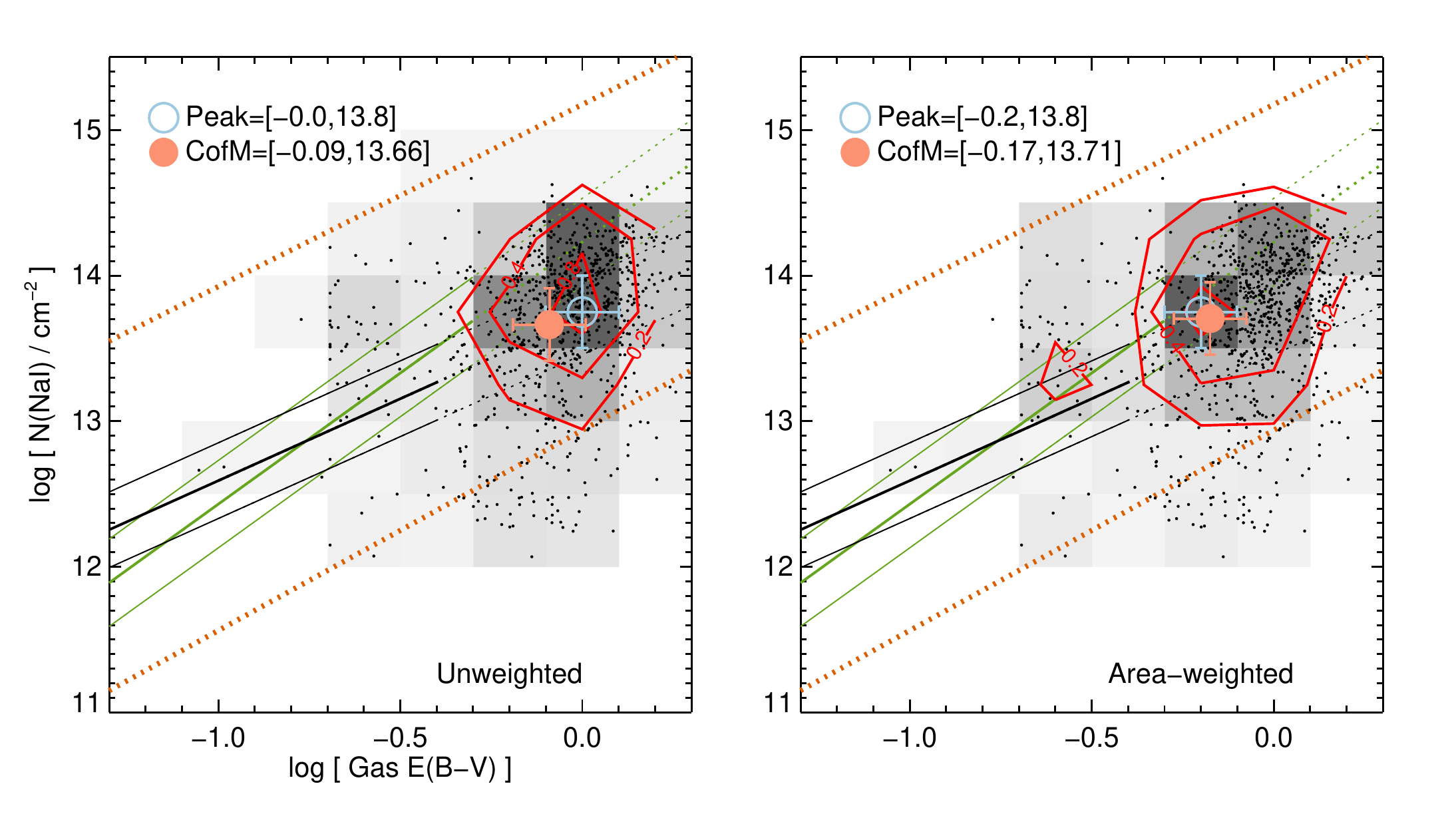}
  \caption{\nad\ absorption column density vs. colour excess from gas measurements in our sample, compared to MW and other external galaxy sightlines. See Fig.~\ref{fig:weq_abs_v_ebv_gas} for more details.}
  \label{fig:nnai_abs_v_ebv_gas}
\end{figure*}

To compare our data to these literature measurements, we take two approaches. First, we consider all unique Voronoi bins in our data and find the 2D histogram of these points in the relevant spaces. We then weight each Voronoi bin by its actual physical area (in kpc$^2$) and recompute the 2D histogram. We also combine the sample to average over the variations seen in Fig.~\ref{fig:weq_v_ebv_stel}--\ref{fig:weq_v_col_gi}. The results are shown in Fig.~\ref{fig:weq_abs_v_ebv_gas} and \ref{fig:nnai_abs_v_ebv_gas}. (The points for individual galaxies, with error bars, are shown in Appendix~\ref{appendixa}.) In these figures we also mark each peak bin and the center-of-mass of each 2D histogram. We treat lower limits in column density the same as other values.

The two weightings and centroid measurements yield almost identical results in $N$(\ion{Na}{1}) vs. \ebv\ space. There is a fairly tight distribution that centers around log$[\ebv]=0.1$ and log($N$/cm$^{-2})=13.7$, with a small tail to lower values in both axes. In \weqa\ vs. \ebv\ space, the distribution centers around $\ebv\sim0.1$, but it is spread out over a wide range of \weq. This is seen in a center-of-mass that is lower in \weq\ than the peak by $\sim$0.2~dex. We note that the area-weighted distribution is centered at lower \weq\ by 0.2~dex and lower \ebv\ by 0.1~dex than the unweighted distribution, probably reflecting that larger-area Voronoi bins tend towards galaxy outskirts and thus lower \weq\ and \ebv.

Our data lie for the most part in the regions delineated by fits to MW
clouds and by the range of values in external galaxies. There are a significant number of low-\weq, low-$N$(\ion{Na}{1}) points with moderate \ebvg\ values which are not present in the comparison data or fits. The reason these points do not arise in MW sightlines or external galaxy sightlines is unclear. It may have to do with the lower covering factors and/or higher \ebvg\ values present in external galaxy sightlines compared to Milky Way sightlines, as well as the remarkable spatial coverage of
external galaxy disks in the current dataset compared to the other
single-aperture external-galaxy probes shown here. The reason that
different galaxies span different regions of these spaces is also
unknown (Appendix~\ref{appendixa}), though we discuss in Section~\ref{sec:discussion} that it may have to do with the different range of linewidths in different systems, and thus potentially the number of clouds.

Despite these differences, the
overall consistency with MW and other external galaxy sightlines is clear and sets a benchmark for future IFS studies of these relationships.

\section{DUSTY INFLOWS AND OUTFLOWS} \label{sec:inflows_and_outflows}

\begin{figure*}
  \includegraphics[width=\textwidth]{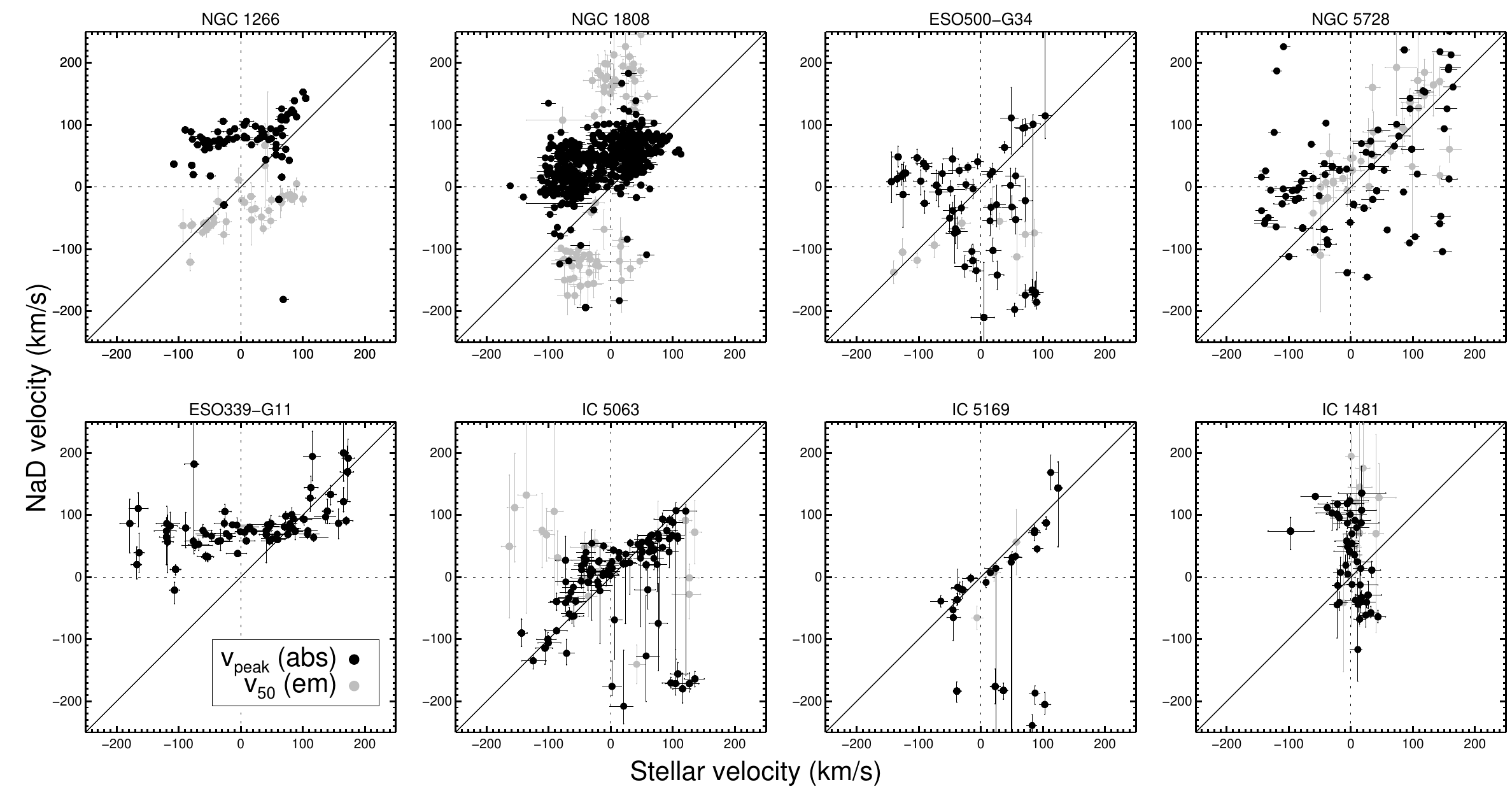}
  \caption{\nad\ velocity ($v_\mathrm{peak}$) vs. stellar
    velocity. \nad\ emission and absorption lines are shown as black
    circles and gray crosses, and error bars are 1$\sigma$. Black
    solid lines of equality and dashed zero-velocity lines are
    overplotted. Clear deviations from stellar rotation are evident in every case.}
  \label{fig:nadvel_v_stelvel}
\end{figure*}

Now that we have established and contextualized the spatially-resolved
connection between cool, neutral gas and dust in these galaxies, we examine
the motions of the neutral gas. We find that the
kinematics of \nad\ absorption and emission of at least some bins in all eight galaxies show significant deviations from the largely circular motions traced by stars (Fig. \ref{fig:nadvel_v_stelvel}).

\begin{figure*}
  \includegraphics[width=\textwidth]{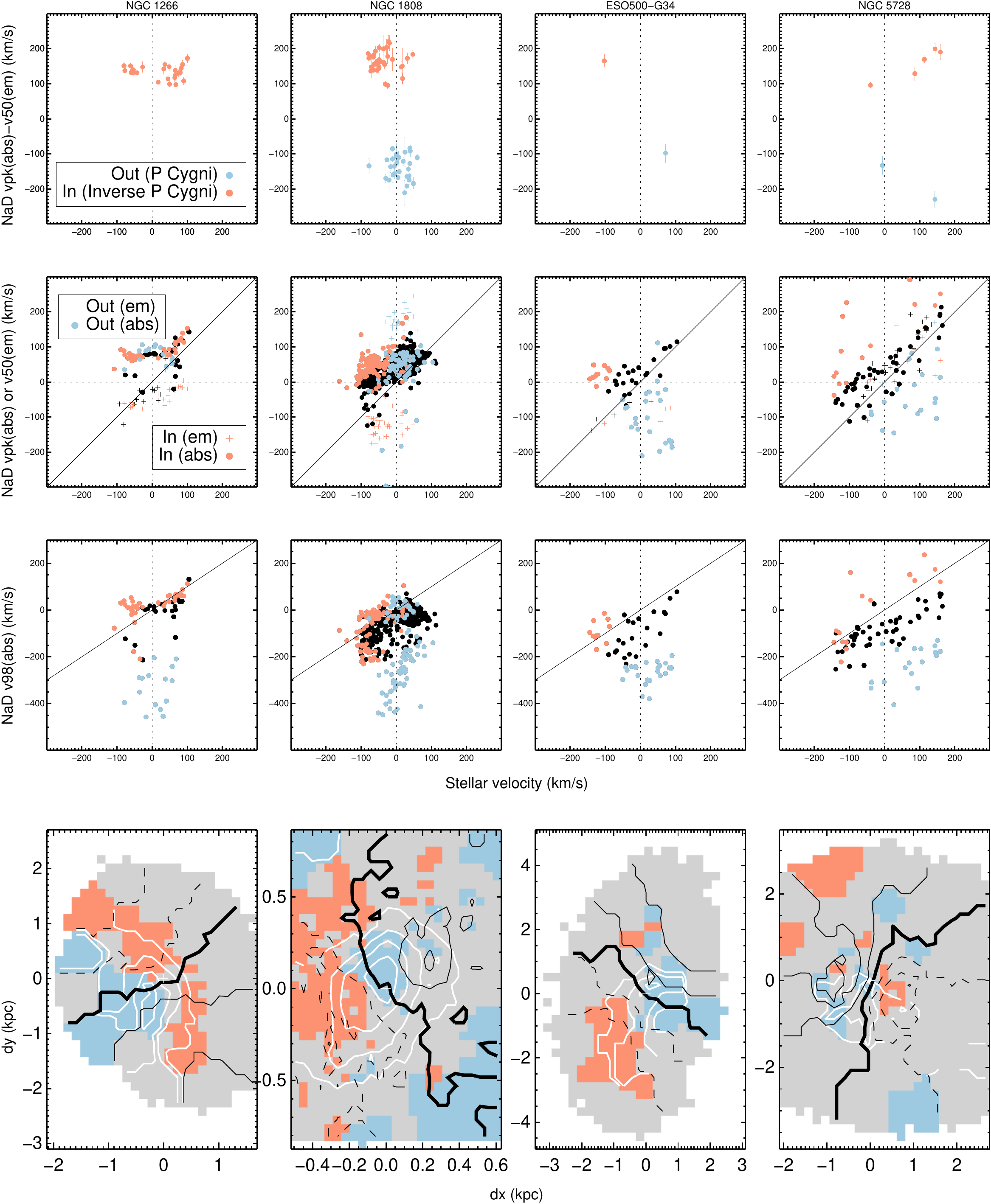}
  \caption{(Top row) Velocity difference between \nad\ absorption and
    emission lines in each Voronoi bin vs. stellar velocity. P-Cygni
    profiles suggesting outflow are negative velocity differences and
    shown in blue; inverse P-Cygni profiles are positive and
    red. 1$\sigma$ errors in velocity difference are shown. (Second row) Peak or central \nad\ velocity vs. stellar
    velocity. Crosses (circles) represent \nad\ emission
    (absorption). Black symbols show bins that do not meet one of
    the criteria for inflow or outflow. Blue (red) symbols show data
    that meet one of the criteria for outflow (inflow); see Section~\ref{sec:inflows_and_outflows}. (Third row) Maximum \nad\
    absorption velocity vs. stellar velocity. (Bottom row) Maps of
    outflow and inflow. Gray shaded regions contain data but no
    detection of outflow or inflow. colour shaded regions are as
    described above. Black contours show stellar velocity; the solid
    thick contour locates $v = 0$~\kms, and the other contours are
    labeled in Fig.~\ref{fig:maps}. White contours show \weqa; labels
    are in \AA.}
  \label{fig:nadvel_meta}
\end{figure*}
\setcounter{figure}{10}
\begin{figure*}
  \includegraphics[width=\textwidth]{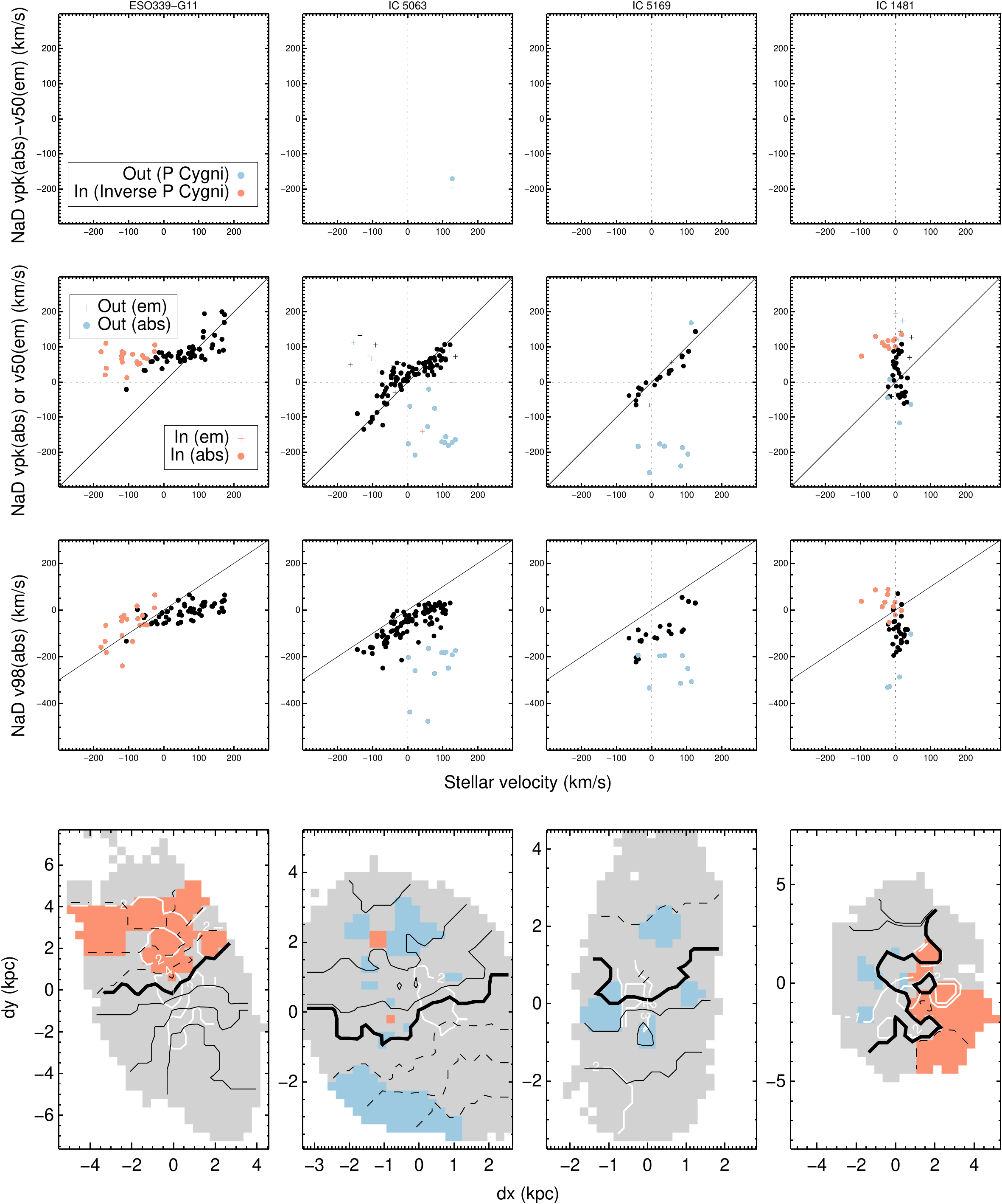}
  \caption{\it Continued.}
\end{figure*}

We use two methods to trace these discrepant motions. The first is by measuring significant velocity differences between \nad\ absorption and emission lines (i.e., P Cygni or
{\it inverse} P Cygni profiles) in each Voronoi bin. This is a typical method for outflow and inflow detection in \nad\ and other resonance lines. The second is to look for significant deviations of either \nad\ absorption or emission from stellar motion. We use as a reference in each Voronoi bin the stellar velocity in that bin. We rely throughout on the expectation from observations \citep{1993AJ....105..486P}, simple models \citep{2011ApJ...728...55R,2019MNRAS.482.4111R}, and radiative transfer calculations \citep{2011ApJ...734...24P} that resonance absorption arises on the near side of a flow (thus showing a redshift during inflow and a blueshift during outflow) while resonance emission arises on the far side of a flow (thus showing a blueshift during inflow and a redshift during outflow). We display these velocity differences for each galaxy in the first three rows of Fig.~\ref{fig:nadvel_meta}.

In the first method, that of P Cygni or inverse P Cygni detection, we require a 2$\sigma$ detection of $v_\mathrm{abs}-v_\mathrm{em} < 0$ for outflow or $v_\mathrm{abs}-v_\mathrm{em} > 0$ for inflow. In practice, almost all of these detections have $|v_\mathrm{abs}-v_\mathrm{em}| > 100$~\kms. Smaller offsets would make the emission and absorption blend into each other. It is possible that blending of these profiles in fact overestimates the velocity offset of these lines in integrated spectra \citep{2011ApJ...734...24P}. For these resolved data, however, emission and absorption resulting from physically distinct locations in an inflow or outflow are expected to have different kinematics, as we discuss above.

In the second method, we require the \nad\ velocity in absorption or emission to differ from the stellar reference velocity by a significant amount. We choose the conservative criteria that the velocities differ from stellar in each Voronoi bin by at least 100~\kms\ (e.g., twice the threshold in \citealt{2005ApJS..160..115R}) and differ from the stellar reference velocity by at least 2$\sigma$. The first criterion means a bin contains inflowing gas when $v_\mathrm{50\%,em}-v_\star<-100$~\kms\ (blueshifted emission)
or $v_\mathrm{50\%,abs}-v_\star>100$~\kms\ (redshifted absorption) and it contains outflowing gas when $v_\mathrm{50\%,em}-v_\star>100$~\kms\ (redshifted emission) or $v_\mathrm{50\%,abs}-v_\star<-100$~\kms\ (blueshifted absorption). The 100~\kms\ threshold may filter out deviations due to radial streaming motions along spiral arms \citep[e.g.,][]{2007ApJ...665.1138S} or bars \citep{1999ApJ...526...97R} or due to tidal motions that appear in merging systems \citep[e.g.,][]{2013ApJ...768...75R}. We also allow an outflow detection if the absorption line wing is highly blueshifted ($v_\mathrm{98\%,abs}-v_\star<-200$~\kms), which is a robust quantity
for defining outflow in starburst galaxies \citep[e.g.,][]{2013ApJ...768...75R}.

These methods are largely consistent with each other; i.e., an inflow or outflow detected with the first method is typically also an outflow as defined using the second method. Examples where the
methods diverge are NGC~5728, in which we observe a few apparently inflowing inverse P Cygni profiles for which the emission component is near the stellar velocity, and NGC~1808, in which we
observe P Cygni profiles for which the absorption component is near
the stellar velocity. In the case of NGC~1808, this is a consequence of absorption tracing the foreground disk in places where the emission traces the outflow \citep{1993AJ....105..486P}; the reverse could be true in NGC~5728.

We detect inflow and/or outflow in every galaxy in our sample over 
projected areas that are 10--40\%\ of the stellar disk visible in the FOV, as shown in the bottom row of Fig.~\ref{fig:nadvel_meta} and in Table~\ref{tab:radarea}. These correspond to physical areas of 1--18~kpc$^2$. We detect outflow in 7/8 systems, and inflow in 7/8 systems.

In 4/7 systems with an outflow (NGC~1266, NGC~1808, ESO 500$-$G34, IC~5169), neutral gas motions are consistent with the picture of a minor-axis flow, roughly oriented along the stellar zero-velocity contour. Two of these, NGC~1266 and NGC~1808, have prior minor-axis outflow detections in \nad\ \citep{2012MNRAS.426.1574D,1993AJ....105..486P}. These four have the highest disk inclinations among the \nad\ outflow detections.

Of the other 3/7 systems with an outflow, IC~5063 has a known radio jet that is accelerating ionized, neutral, and molecular gas along the major axis \citep{1998AJ....115..915M,2007A&A...476..735M,2015A&A...580A...1M}. The outflowing gas we detect in \nad\ is likely the dusty counterpart. In NGC~5728, there is a known ionized, minor-axis outflow \citep{2019ApJ...870...37D,2019ApJ...881..147S,2019MNRAS.490.5860S} that is blueshifted in the NW and redshifted in the SE (corresponding to the bottom and top of our IFS maps). Some of the motions we observe may be connected with this narrow line region outflow, but not all Voronoi bins with detected outflow are consistent with this orientation. Finally, IC~1481 has an irregular morphology, low inclination, and twisted stellar isovelocity contours. Thus, determining the minor axis is difficult and the relationship between the gas kinematics and galaxy structure is unclear.

The inflow signatures we observe are spatially coherent in most cases. While the relationship between these regions and the underlying galaxy structure is not obvious in every galaxy, the inflow often lies closer to the projected major axis than the minor axis of the system.  In most of these cases, the spatial footprint of the inflowing gas is largely confined to one side of the disk, with the exceptions of NGC~1266 and NGC~5728. 4/7 of the systems with inflows contain bars, and all are spirals.

\nad\ emission is present in almost every galaxy in the sample, and P Cyngi or inverse P Cygni profiles show up in 5/8 systems. P Cygni profiles (blueshifted absorption and redshifted emission) are observed in spatially-resolved observations of galactic outflows in a few nearby galaxies \citep{1993AJ....105..486P,2015ApJ...801..126R,2019AA...623A.171P,2020MNRAS.494.5396B}, as well as others at high redshift \citep{2011ApJ...728...55R,2013ApJ...770...41M,2016MNRAS.458.1891B}, and are a result of radiation transfer effects in the outflow \citep{2011ApJ...734...24P}.  However, to our knowledge, this is the first detection of {\it inverse} P Cygni profiles across galactic disks.

Besides the orientation of the flows, the other properties of the outflowing and inflowing gas are remarkably similar. As shown in Table~\ref{tab:radarea}, both flows typically extend to 50--75\%\ of the edge of the stellar disk detected by WiFeS, as seen by comparing the maximum observed radii of the disk $R_\mathrm{max}^\mathrm{disk}$ to the maximum observed inflow and outflow radii, $R_\mathrm{max}^\mathrm{out}$ and $R_\mathrm{max}^\mathrm{out}$. The maximum inflow and outflow radii
range from 1 to 6 kpc. They also are projected onto a similar areal percentage of the stellar disk, up to 25\%\ (median 18\%\ for inflows and 12\%\ for outflows). The higher median value for the inflows may reflect the tendency for these flows to be closer to the major axis than to the minor axis.

There are not systematic differences across the sample in log($N($\ion{Na}{1}$)$/cm$^{-2}$) or \weq\ between the inflow and outflow regions. The area-weighted average and standard deviation of log($N($\ion{Na}{1}$)$/cm$^{-2}$) are (13.6$\pm$0.6) and (13.5$\pm$0.6) for inflow and outflow, respectively. For \weq, the average and standard deviation are (1.9$\pm$1.8)~\AA\ and (1.6$\pm$2.9) for inflow and outflow. The luminosity- and area-weighted statistics are comparable.

\begin{table}
    \centering
    \caption{Projected radii and areas, as displayed in the bottom row of Fig.~\ref{fig:nadvel_meta}. Listed are the projected maximum radius of the disk $R_\mathrm{max}^\mathrm{disk}$, inflow $R_\mathrm{max}^\mathrm{in}$, and outflow $R_\mathrm{max}^\mathrm{out}$; the projected disk area $A^\mathrm{disk}$; and the percent of projected disk area classified as
    showing inflow, $A^\mathrm{in}$, or outflow, $A^\mathrm{out}$.}
    \label{tab:radarea}
    \begin{tabular*}{\columnwidth}{c @{\extracolsep{\fill}} ccccrr}
        \hline
        Galaxy & $R_\mathrm{max}^\mathrm{disk}$ & $R_\mathrm{max}^\mathrm{in}$ & $R_\mathrm{max}^\mathrm{out}$ & $A^\mathrm{disk}$ & $A^\mathrm{in}$ & $A^\mathrm{out}$ \\
        --- & kpc & kpc & kpc & kpc$^2$ & \% & \% \\[0.1cm]
        \hline
  NGC 1266&  2.54&  2.14&  1.97& 13.45&    25&    17\\
  NGC 1808&  1.04&  0.86&  1.04&  1.94&    18&    25\\
ESO500-G34&  4.54&  3.29&  2.65& 36.14&    13&    12\\
  NGC 5728&  4.04&  3.18&  3.72& 30.16&     7&     9\\
ESO339-G11&  8.53&  6.24&  0.00& 85.89&    20&     0\\
   IC 5063&  4.56&  3.22&  3.74& 39.29&     1&    17\\
   IC 5169&  4.55&  0.00&  2.49& 25.38&     0&    10\\
   IC 1481&  5.59&  5.44&  2.50& 62.84&    24&     4\\[0.1cm]
        \hline
    \end{tabular*}
\end{table}

\section{DISCUSSION} \label{sec:discussion}

We have detected ubiquitous and spatially-extended flows of cool, neutral gas across our S7 subsample that differ significantly from stellar motion. The correlations between gas and dust properties point to the dusty nature of these flows. We have characterized these correlations within individual galaxies and within the sample as a whole to compare to previous results and to benchmark future, larger integral-field studies of external galaxies.

The correlations between $N$(\nad) or \weq\ and \ebv\ have not been extensively characterized in external galaxies, particularly at the spatially-resolved level. Thus our subsample, despite its modest size in terms of number of galaxies, represents a significant addition. The relevant observables have for decades been used to characterize the Milky Way's ISM, so connecting these data to other galaxies is an important step.

In particular, we show that the relationship between \weq\ and various dust measures is not the same from galaxy to galaxy (Fig.~\ref{fig:weq_v_ebv_stel}--\ref{fig:weq_v_col_gi}). The correlation between \weq\ and measures of colour excess probably reflects a 3D spatial connection between neutral gas and dust. However, \weq\ and \ebv\ are indirect measures of the amount of gas and dust along a given line of sight. In these data, absorption lines dominate total \weq\ in most spaxels, with resonant line emission biased toward low-\ebv\ sightlines. Absorption-line \weq\ is directly connected to column density (the curve of growth), but it is modulated by covering factor. A high $N$(\ion{Na}{1}) line of sight with low $C_f$ can have the same \weq\ as a line with low $N$(\ion{Na}{1}) and high $C_f$. $N$(\ion{Na}{1}) is in turn dependent on optical depth and linewidth.

\begin{figure*}
  \includegraphics[width=\textwidth]{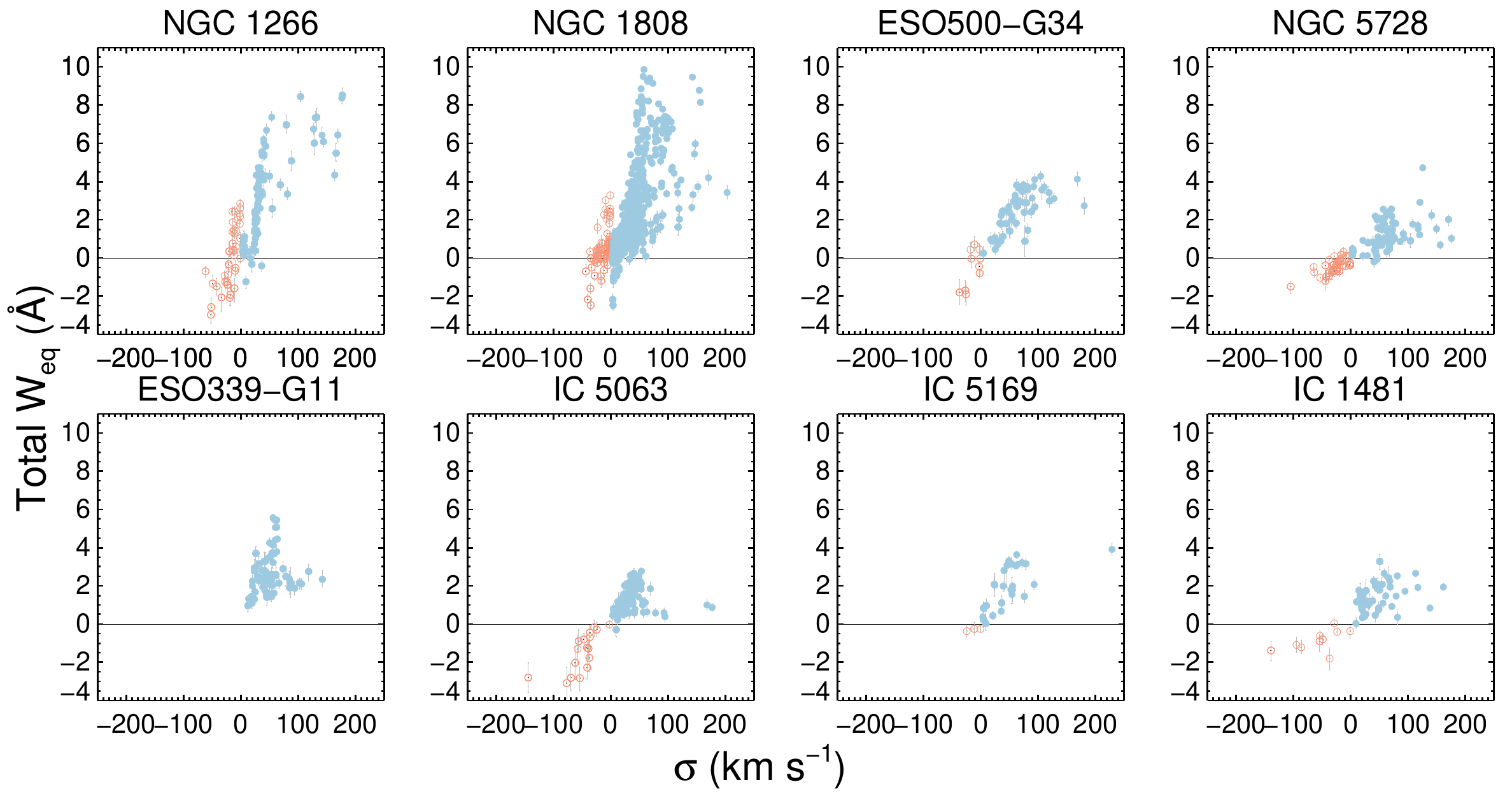}
  \caption{Total fitted \nad\ equivalent width vs. velocity dispersion
    in each component. Spaxels with more than one component are
    represented by multiple points. Absorption components are shown in
    light blue with positive $\sigma$; emission components are shown in light red
    with negative $\sigma$. The median and standard deviation of correlation coefficients are 0.6$\pm$0.2 for absorption and 0.7$\pm$0.2 for emission. This illustrates that $\sigma$ is one of the primary parameters modulating equivalent width.}
  \label{fig:weq_v_sig}
\end{figure*}

Based on nuclear galaxy spectra, \citet{2000ApJS..129..493H} find that
the depth of the \nad\ profile (which they suggest depends primarily
on $C_f$) is correlated with stellar colour and colour excess from gas measurements. They
argued that increasing dust obscuration is thus directly connected to
the increasing covering factor of the neutral gas. However, we do not
find that equivalent width correlates strongly with either $\tau$ or $C_f$. The linear Pearson correlation coefficients of \weqa\ with \cf\ and $\tau$ are $-0.1\pm0.3$ and $0.2\pm0.3$ (median plus or minus standard deviation). Instead, we find that total \weq\ increases with
increasing absorption linewidth and decreases with increasing emission
linewidth in most systems (Fig.~\ref{fig:weq_v_sig}). The linear Pearson correlation coefficients of \weq\ with $\sigma_\mathrm{abs}$ and $\sigma_\mathrm{em}$ are $0.6\pm0.2$ and $-0.7\pm0.2$. Thus, equivalent
width, colour excess produced by dust, and total line dispersion
are intimately correlated. If the correct picture is the collection of
compact clouds discussed above (Section~\ref{sec:results}; \citealt{2019ApJ...887...89W,2019ApJ...886L..13P}), then these quantities may each scale with the number of clouds along the line of sight.

Considering all of the points within our sample, gas and dust correlations extend relationships determined from lower-\ebv\ sightlines within the Milky Way and some external sightlines. In $N$(\nad) vs. \ebv\ space, they straddle correlations from \citet{1974ApJ...191..381H} and \citet{2013ApJ...779...38P} and lie in the same space as high-\ebv\ infrared luminous galaxies. In \weq\ vs. \ebv\ space, they are on average very consistent with relationships from low-$z$ star-forming galaxies. They also largely lie within the envelope of previous single-sightline observations of infrared-luminous and star-forming galaxies at high \ebv\ \citep{2005ApJS..160...87R,2008ApJ...674..172R,2010AJ....140..445C}, though with more scatter to lower \weq\ due to the sensitivity of our data. This in turn suggests that the dusty, neutral gas in these flows likely has similar physical conditions to gas in the disks and halos of the Milky Way and nearby star-forming galaxies. These similarities have been used to, e.g., calculate \ion{H}{1} column densities from \nad\ properties \citep[e.g.,][]{2005ApJS..160...87R}. Thus, our results lend strength to the \ion{H}{1} column densities estimated in this way, despite the uncertainties due to ionization state and dust depletion. The mass outflow rates computed for neutral outflows---which are critical for evaluating their role as negative feedback in galaxy evolution---in turn depend on these column density estimates \citep[e.g.,][]{2005ApJS..160..115R}.

We do not claim that this sample is at all representative, either of the S7 sample as a whole or of local galaxies more generally. In particular, the subsample has been selected for its strong \nad\ signatures. However, it is well-suited for characterizing the \nad\ feature and its connection to dust, as well as illustrating how it can be used to detect both outflows and inflows. We also note that the signatures of these phenomena are widely-distributed across the disks of these systems and would not be well-characterized by single-aperture spectra. Thus integral-field studies of these flows, though their sample sizes are inherently limited, are critical to determining their properties and impact.

The detection of inverse P Cygni profiles in \nad\ in half of these data cubes is potentially a first for external galaxy sightlines. In several of these data cubes, the signature is seen in a large number of spaxels. Sensitive, spatially-resolved observations help to uncover these, since the nuclear spectra tend to be dominated by absorption or outflow. Higher spectral resolution is key, as well ($R=7000$ at \nad\ for WiFeS), since the linewidths in spaxels showing these inverse P Cygni profiles are narrow (Fig.~\ref{fig:inversePCygni}) and the velocity shifts are not large ($\la$200~\kms; Fig.~\ref{fig:nadvel_meta}). Previous single-aperture studies like those with the  SDSS ($R\sim2000$ at \nad) significantly smear and weaken these signatures, making them more difficult to interpret. Finally, moderate-resolution spectral templates \citep{2004ApJS..152..251V} yield high-quality subtraction of the stellar continuum.

Inverse P Cygni profiles in \nad\ are seen in spectra of T Tauri and Herbig Ae stars \citep{1994AJ....108.1056E,1994A&A...292..165G}, in which the signature is interpreted as accretion. The picture sketched by \citet{2019MNRAS.482.4111R} in which inverse P Cygni profiles should appear in lines of sight through high-inclination disks due to infall appears to apply here. The inflow signatures we observe prefer the projected major axis and 7/8 of our disks have $i>50^\circ$. Presumably the same radiation transfer effects that arise in the outflow and produce P Cygni profiles can produce their inverse. Radiative transfer simulations, like those in \citet{2011ApJ...734...24P}, could test this postulate.

Despite the selection bias, the frequent detection of radial motions consistent with outflow and inflow is striking. Our results are thus of a piece with the ubiquity of both inflow and outflow signatures in nearby AGN from single-aperture studies \citep{2010ApJ...708.1145K,2019MNRAS.482.4111R,2019MNRAS.486.1608N}. Single-aperture studies have also found that inflow dominates in edge-on systems and outflow in face-on systems \citep[e.g.,][]{2010AJ....140..445C,2019MNRAS.482.4111R}. A similar result in our spatially-resolved study is the alignment of outflow signatures along the minor axis and the signatures of inflow that lie closer to the projected major axis. We find both signatures in many cases, rather than inflow or outflow dominating, despite 7/8 galaxies having inclination $i>50^\circ$.

While the disks in these galaxies may collimate an inner wind regardless of the power source, we note that NGC 5728 \citep{2018ApJ...867..149D}, IC~5063 \citep{2007A&A...476..735M}, and possibly NGC~1266 \citep{2013ApJ...779..173N} contain small-scale radio jets. In NGC~5728 (and perhaps NGC~1266), this jet is oriented roughly along the projected minor axis, but in IC~5063 it is oriented along the projected major axis. In IC~5063, we observe that the outflowing neutral gas also preferentially lies along the major axis (Fig.~\ref{fig:nadvel_meta}).

The inflows we observe can be caused by gravitational effects in the disks themselves, such as bars, spiral arms, or other non-axisymmetric structures. We choose a conservative criterion for defining inflow that may preclude these possibilities in some cases, but we cannot rule them out. NGC~1808 has radial streaming motions that are reflected in ionized and molecular gas \citep{2016ApJ...823...68S, 2017A&A...598A..55B}. It also has a steeper gas rotation curve than the stars. The molecular gas in IC~5063 shows a steep rotation curve, as well \citep{2015A&A...580A...1M}. Finally, NGC~5728 hosts a stellar bar, nuclear star-forming ring, and spiral arms, all of which produce gas dynamics different from the stars \citep{2019MNRAS.490.5860S}.

Undoubtedly there is some contribution from non-axisymmetric potentials to the neutral gas motions we observe. Tidal motions may also contribute in, e.g., IC~1481. However, the magnitudes of the discrepant ionized and molecular gas motions in NGC~1808, IC~5063, and NGC~5728 (of the order $\sim$100~\kms\ or less compared to stellar rotation) are below our detection threshold if these motions were also present in neutral gas traced by \nad. Furthermore, if we choose the ionized gas \vfifty\ or the velocity at peak flux in place of stellar velocity as a reference point, we find even more widespread detection of inflow and outflow in these systems. These suggest that the motions we observe may not be dominated by inflow due to secular disk structures.

A plausible alternative is that the dusty inflows we observe at projected radii $<$10~kpc are the endpoint of inflows of cold, circumgalactic gas. Co-rotating gas is observed in \ion{Mg}{2} absorption in extended galactic disks at a substantial fraction of the virial radius (i.e., well into the halo or circumgalactic medium; \citealt{2017ApJ...835..267H,2019MNRAS.485.1961Z}). This gas may be radially inflowing at modest velocities ($\la$60~\kms; \citealt{2019ApJ...875...54H}). The inflow velocities we observe are typically higher at 100--200~\kms, though we intentionally exclude some inflowing gas at lower velocities and are limited in sensitivity to very low-velocity gas. This high-velocity gas could represent an acceleration of these circumgalactic inflows at small radius as the gas falls down the gravitational well.

\section{SUMMARY} \label{sec:summary}

Our understanding of dusty, neutral flows in galaxies is increasingly being informed by integral-field spectroscopy of external galaxies. We here present a new analysis of eight data cubes from the S7 survey, a large IFS survey of nearby Seyfert galaxies in the Southern hemisphere. Uniquely, these data sensitively cover a wide field of view at a relatively high spectral resolution for galaxy surveys ($R=7000$ at \nad). These galaxies were selected for their widespread \nad\ signatures, including both resonant emission and absorption.

These data enable us to characterize the spatially-resolved relationship between the cool, neutral gas (as traced by \nad) and dust columns (as traced by stellar and gas \ebv) in these systems. While this relationship has been studied extensively in the Milky Way, it is relatively unconstrained in external galaxy sightlines and detailed spatially-resolved studies exist for only a few galaxies. We find that these eight disks are consistent with previous measurements and extrapolations at $\ebv\sim1$, though the range of \weqa\ and $N$(\nad) traced by individual spatial locations is much larger, particularly in \weqa. Milky Way measurements of \nad\ and \ion{H}{1} have been used to estimate the mass and energetics of galactic winds from \nad\ only \citep[e.g.,][]{2005ApJS..160...87R}. The consistency we measure among Milky Way and external galaxy measurements of \nad, \weq, and \ebv\ lends credence to these estimates.

Within individual galaxies, strong and significant correlations exist between total \weq\ and \ebvs, \ebvg, and/or optical colour $g-i$. The correlations of \weq\ with $g-i$ and \ebvs\ are the strongest. The parameters of linear regressions are on average consistent with physical expectations for \ebvs/\ebvg\ and for intrinsic unattenuated colours. However, the linear slopes vary from galaxy to galaxy, and in half of the sample the three dust measures show some differences in how they relate to \weq. We find that equivalent width differences within a galaxy are largely driven by velocity dispersion. If compact clouds with small dispersion make up the cool ISM of these galaxies, then the correlations between \nad\ and colour excess may reflect a changing number of clouds along the line of sight.

Using Doppler shifts of both resonant emission and absorption, we find spatially coherent signatures of high-velocity  non-rotational flows ($|v_\mathrm{abs}-v_\mathrm{em}| > 100$~\kms\ and/or $|v_\mathrm{abs/em}-v_\star| > 100$~\kms) in each of these high-inclination galaxy disks (7/8 have $i>50^\circ$). Outflow is observed in both blueshifted absorption or redshifted emission (or perhaps both; a P Cygni profile), while inflow is observed in redshifted absorption or blueshifted emission (or both; an inverse P Cygni profile). This is consistent with recent work on large, single-aperture surveys of nearby AGN \citep{2010ApJ...708.1145K,2019MNRAS.482.4111R,2019MNRAS.486.1608N}.  We find that the column density, equivalent width, size, and area subtended by the inflows and outflows are not substantially different on average. The outflows are consistent with minor-axis collimation and/or jet driving, while the inflows are more closely aligned with the major axes. While some of the inflow motions may reflect tidal motions or secular streaming motions due to, e.g., bar or spiral arm potentials, their relatively high velocities could also point to other mechanisms such as the endpoints of halo-scale accretion.

For perhaps the first time in external galaxies, we detect ubiquitous inverse P Cygni profiles in \nad, which is enabled by the relatively high spectral resolution of the data and spectral libraries of correspondingly high resolution. Profiles like this are found in accreting Galactic T Tauri and Herbig Ae stars \citep{1994AJ....108.1056E,1994A&A...292..165G}, and in these cases are presumably due to infall onto galactic disks \citep{2019MNRAS.482.4111R}.

Ongoing and future IFS surveys like MaNGA and the Hector Galaxy Survey will probe these relationships over much larger sample sizes. The present sample exemplifies the importance of IFS for probing the nature of neutral gas flows in external galaxies, suggesting that these IFS surveys are necessary for improving on single-aperture studies of the cool, neutral medium. The current sample thus serves as a bridge between studies of the Milky Way and these larger IFS surveys, which will be able to better characterize the full parameter space of \nad\ strength and dust column and galaxy type.

\section*{ACKNOWLEDGMENTS}

D.S.N.R. and A.D.T. dedicate this paper to the memory of their co-author, Mike Dopita. Mike initiated and inspired this project with his usual infectious enthusiasm. D.S.N.R. thanks Lisa Kewley and the GEARS3D group at Australian National University for their gracious hospitality while this work was begun. The authors thank Zhao-Yu Li and Luis Ho for providing images from the CGS survey and Mark Phillips for providing supernova \nad\ measurements. Finally, the authors thank the referee for a thorough and detailed report that significantly improved the manuscript. D.S.N.R. was supported in part by the J. Lester Crain Chair of Physics at Rhodes College and by a Distinguished Visitor grant from the
Research School of Astronomy \&\ Astrophysics at Australian National
University.  A.D.T. acknowledges the support of the Australian Research Council through Discovery Project \#DP160103631.

This research has made use of the NASA/IPAC Extragalactic Database (NED), which is funded by NASA and operated by Caltech. 
 
We acknowledge use of the HyperLeda database (http://leda.univ-lyon1.fr).

The Pan-STARRS1 Surveys (PS1) and the PS1 public science archive have
been made possible through contributions by the Institute for
Astronomy, the University of Hawaii, the Pan-STARRS Project Office,
the Max-Planck Society and its participating institutes, the Max
Planck Institute for Astronomy, Heidelberg and the Max Planck
Institute for Extraterrestrial Physics, Garching, The Johns Hopkins
University, Durham University, the University of Edinburgh, the
Queen's University Belfast, the Harvard-Smithsonian Center for
Astrophysics, the Las Cumbres Observatory Global Telescope Network
Incorporated, the National Central University of Taiwan, the Space
Telescope Science Institute, NASA under Grant No. NNX08AR22G issued through the Planetary Science Division of the NASA Science Mission Directorate, the National
Science Foundation Grant No. AST-1238877, the University of Maryland,
Eotvos Lorand University (ELTE), the Los Alamos National Laboratory,
and the Gordon and Betty Moore Foundation.

The VST ATLAS data products are from observations made with ESO
Telescopes at the La Silla Paranal Observatory under program ID
177.A-3011(A,B,C,D,E,F,G,H,I,J) \citep{2015MNRAS.451.4238S}.

The national facility capability for SkyMapper has been funded through
ARC LIEF grant LE130100104 from the Australian Research Council,
awarded to the University of Sydney, the Australian National
University, Swinburne University of Technology, the University of
Queensland, the University of Western Australia, the University of
Melbourne, Curtin University of Technology, Monash University and the
Australian Astronomical Observatory. SkyMapper is owned and operated
by The Australian National University's Research School of Astronomy
and Astrophysics. The survey data were processed and provided by the
SkyMapper Team at ANU. The SkyMapper node of the All-Sky Virtual
Observatory (ASVO) is hosted at the National Computational
Infrastructure (NCI). Development and support the SkyMapper node of
the ASVO has been funded in part by Astronomy Australia Limited (AAL)
and the Australian Government through the Commonwealth's Education
Investment Fund (EIF) and National Collaborative Research
Infrastructure Strategy (NCRIS), particularly the National eResearch
Collaboration Tools and Resources (NeCTAR) and the Australian National
Data Service Projects (ANDS).

\bibliographystyle{mnras}
\bibliography{s7nad}

\section*{DATA AVAILABILITY}

The data underlying this paper, and fits to the data, are available online at \url{https://cloudstor.aarnet.edu.au/plus/s/zyE9jAjHWcGb12v} or \url{https://docs.datacentral.org.au/s7/}. Spectral fitting software used in the paper is available at \url{https://github.com/drupke/ifsfit}.


\section*{SUPPORTING INFORMATION}

Additional Supporting Information may be found in the online version of this article:\\

{\bf Figure~\ref{fig:maps}}. Images and maps of stellar and gas properties of each galaxy.\\

\setcounter{figure}{0}
\begin{figure*}
 \includegraphics[width=\textwidth]{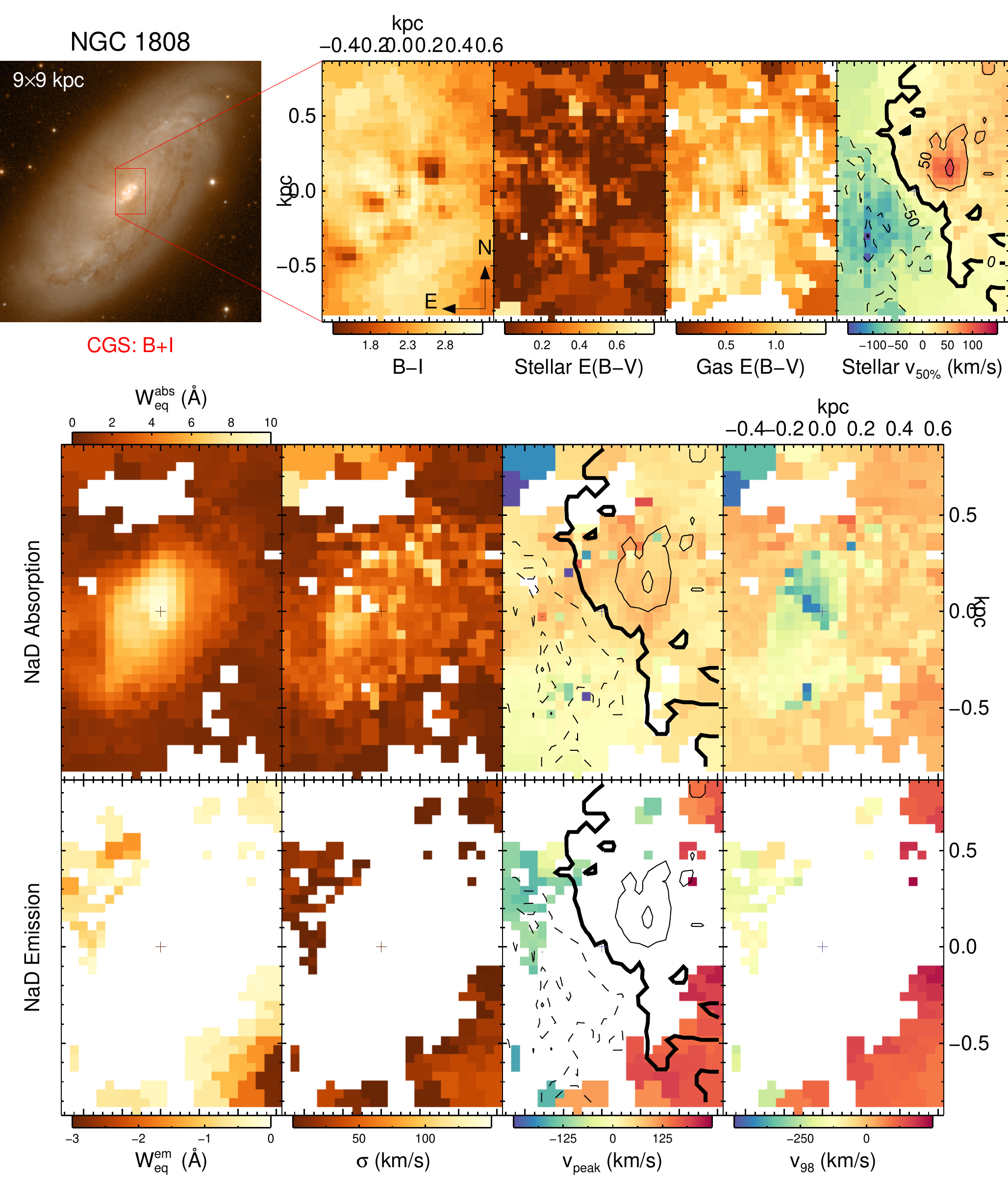}
 \caption{\it Continued.}
\end{figure*}
\setcounter{figure}{0}
\begin{figure*}
 \includegraphics[width=\textwidth]{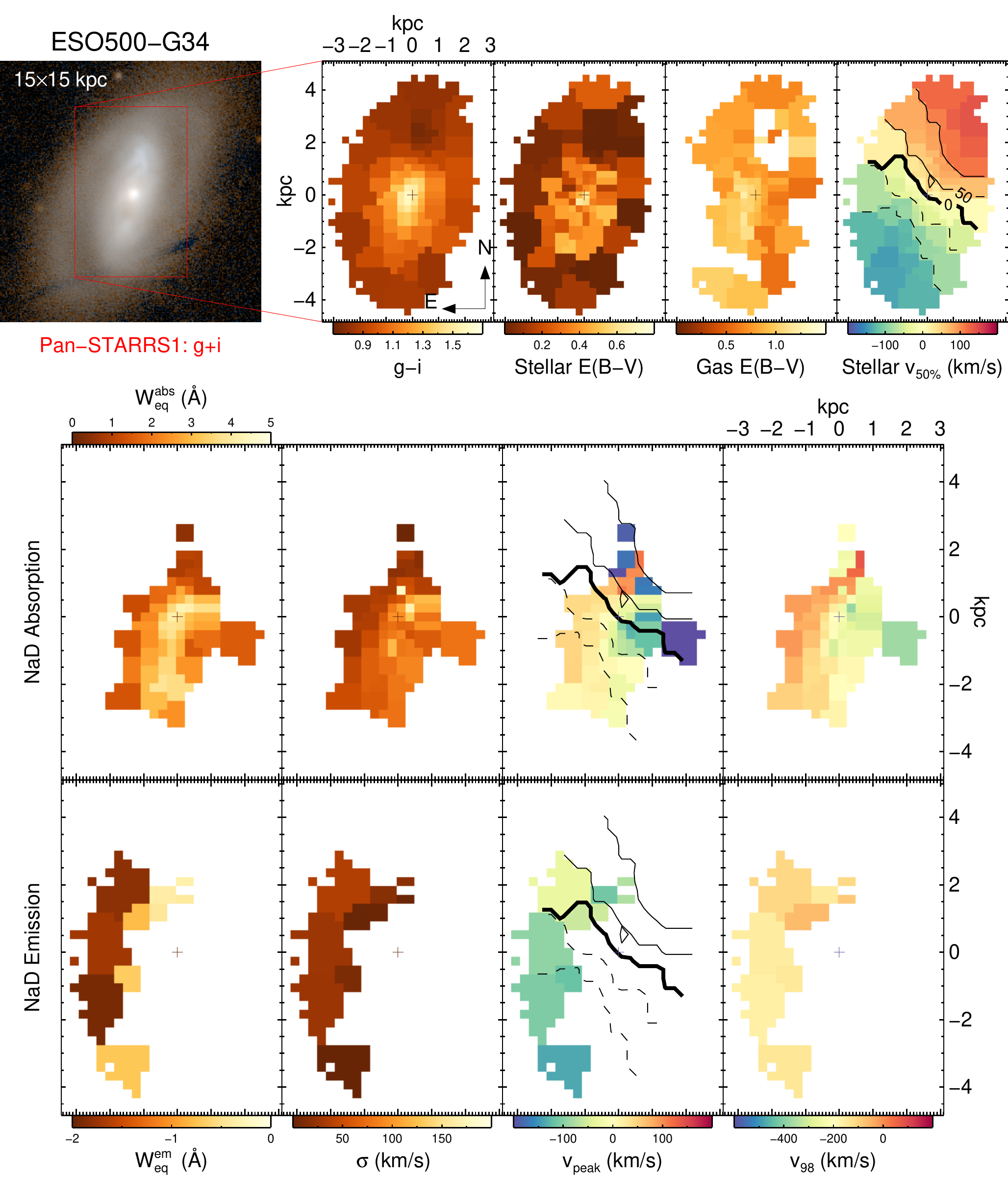}
 \caption{\it Continued.}
\end{figure*}
\setcounter{figure}{0}
\begin{figure*}
 \includegraphics[width=\textwidth]{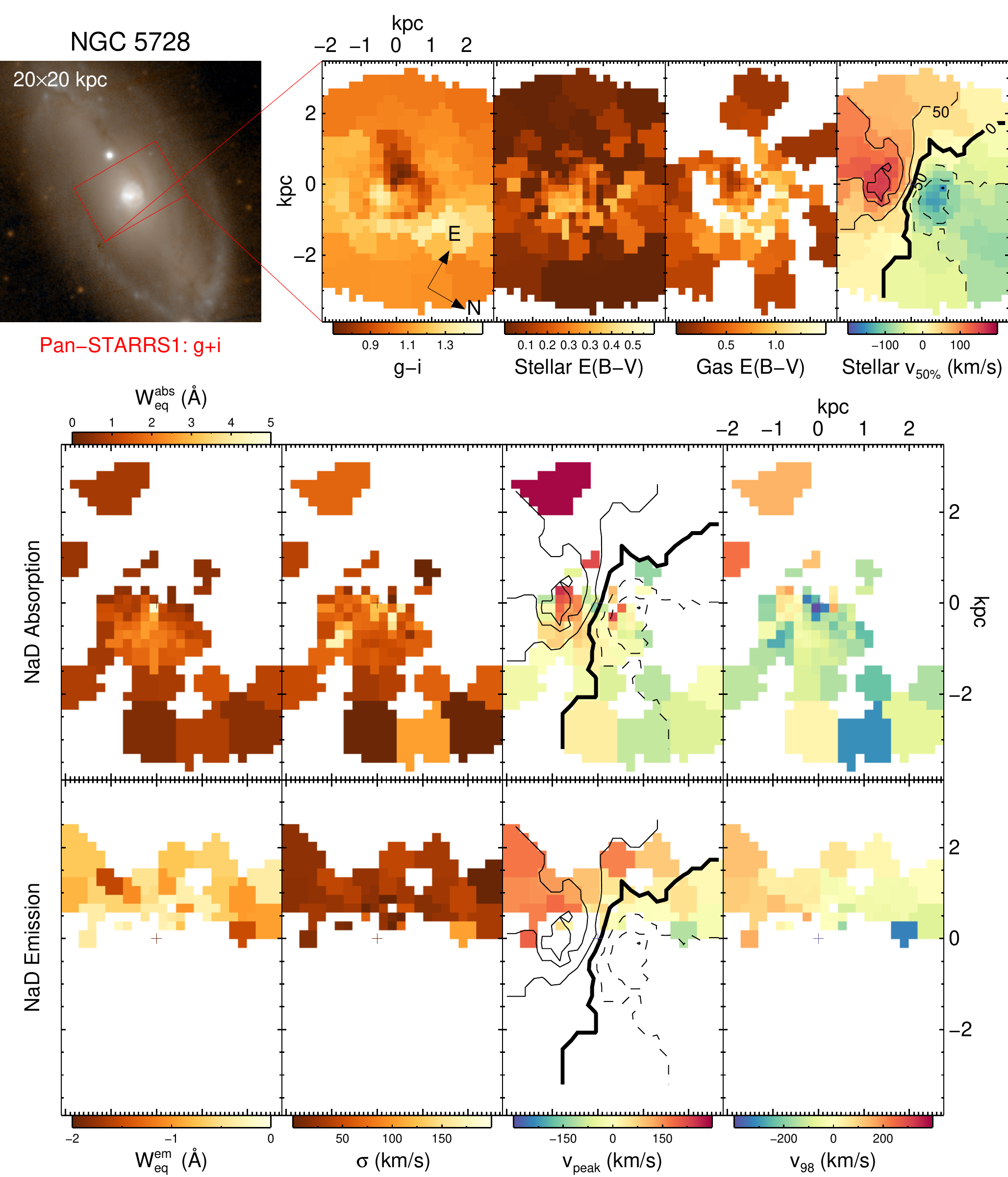}
 \caption{\it Continued.}
\end{figure*}
\setcounter{figure}{0}
\begin{figure*}
 \includegraphics[width=\textwidth]{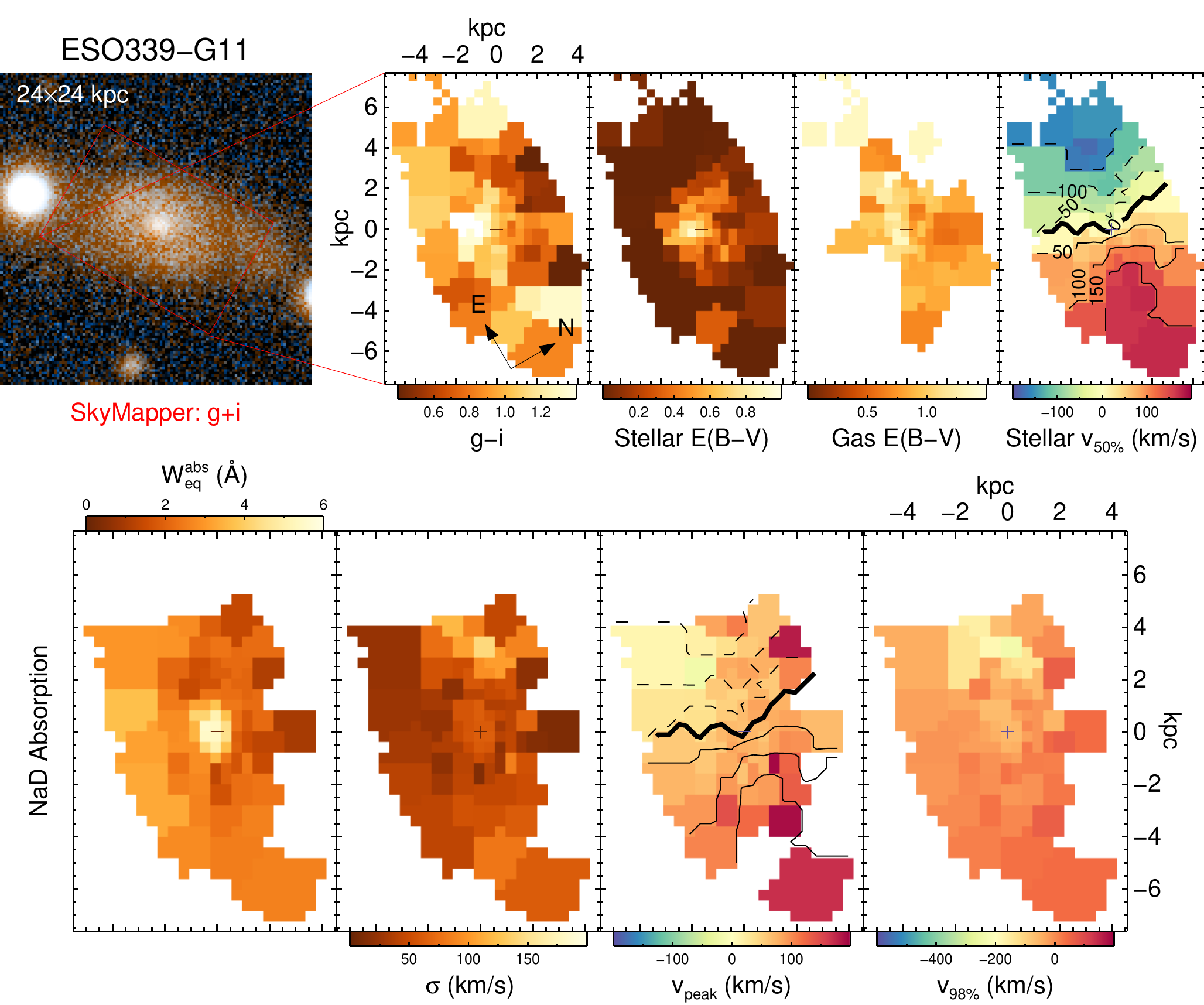}
 \caption{\it Continued.}
\end{figure*}
\setcounter{figure}{0}
\begin{figure*}
 \includegraphics[width=\textwidth]{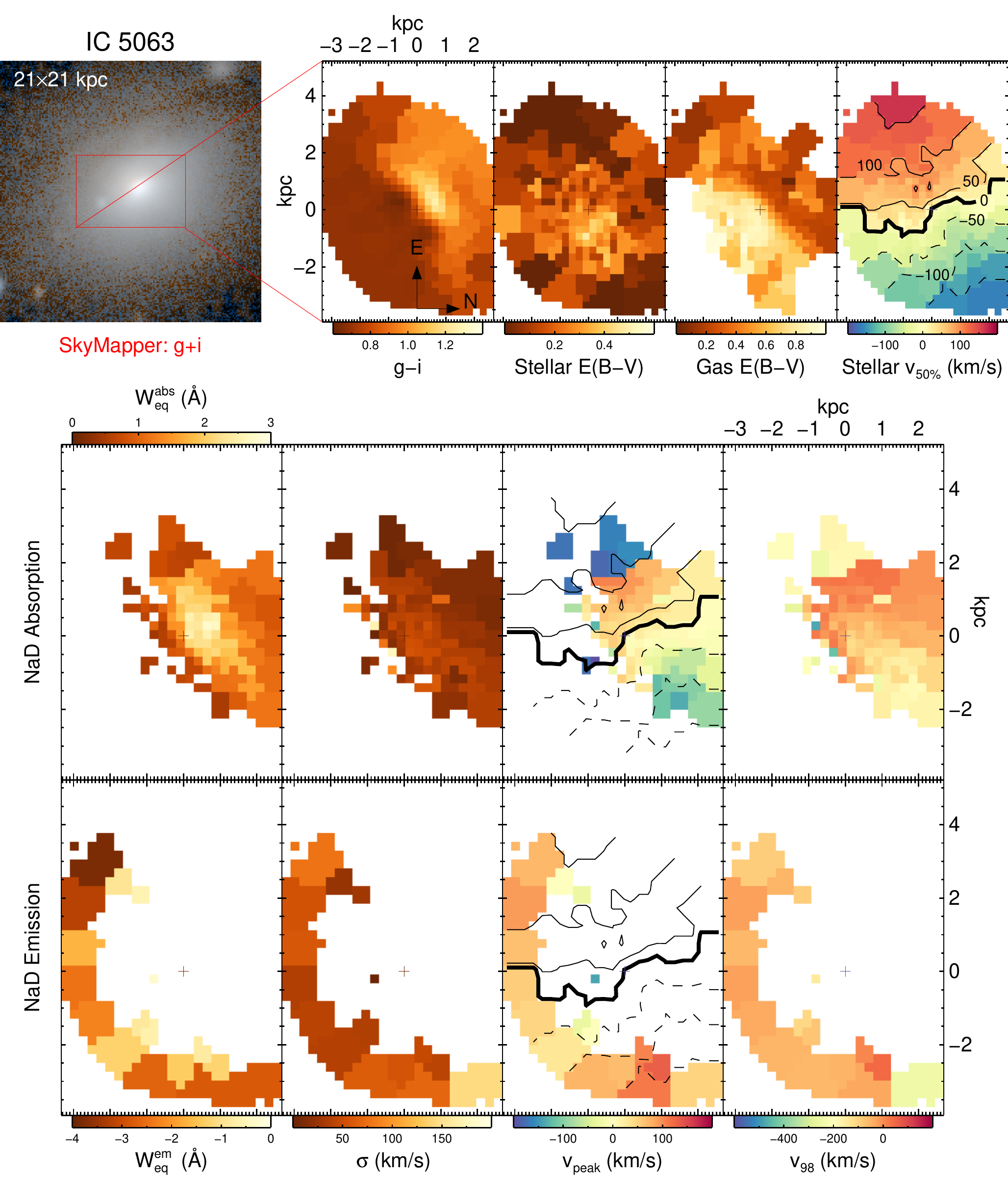}
 \caption{\it Continued.}
\end{figure*}
\setcounter{figure}{0}
\begin{figure*}
 \includegraphics[width=\textwidth]{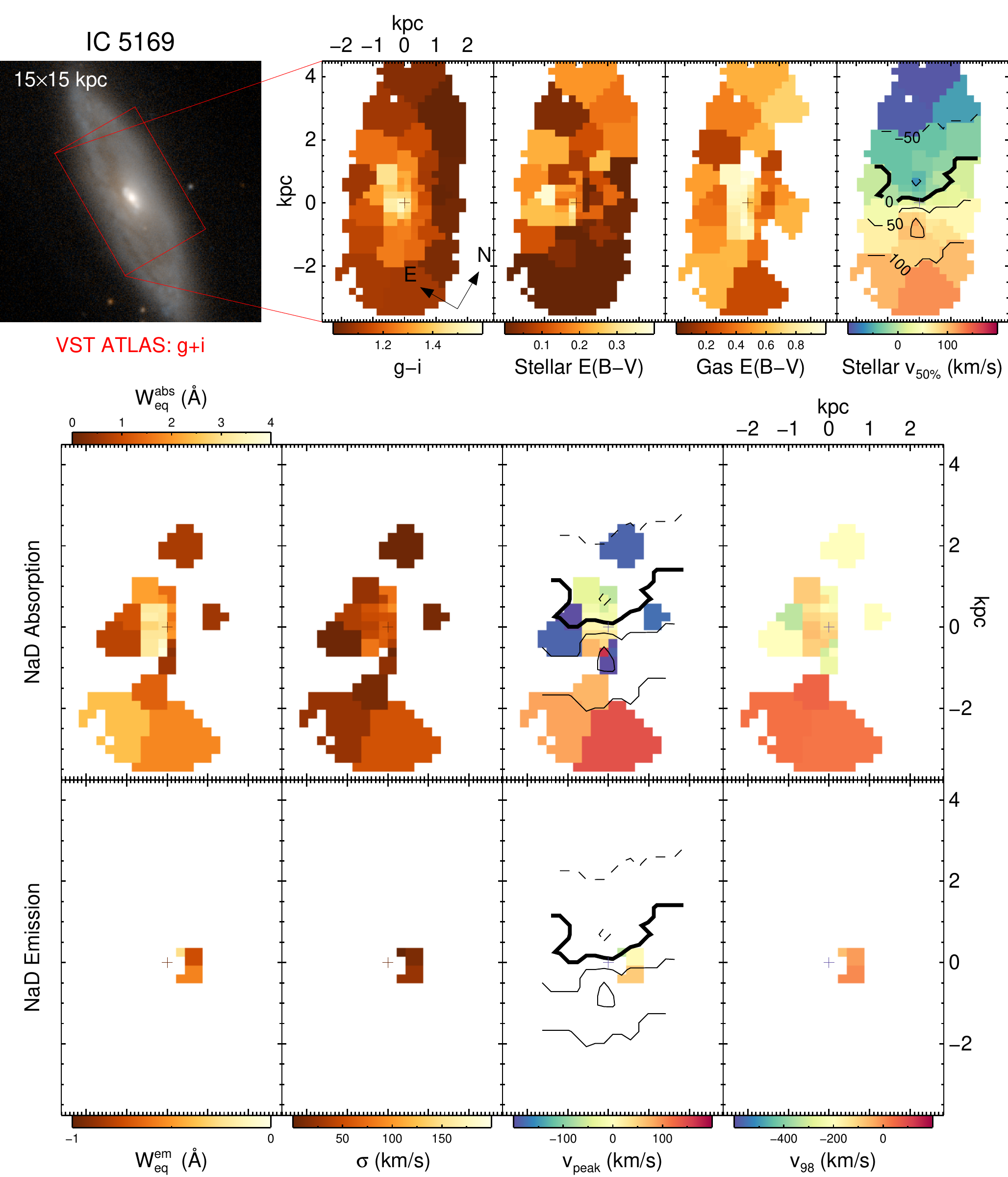}
 \caption{\it Continued.}
\end{figure*}
\setcounter{figure}{0}
\begin{figure*}
 \includegraphics[width=\textwidth]{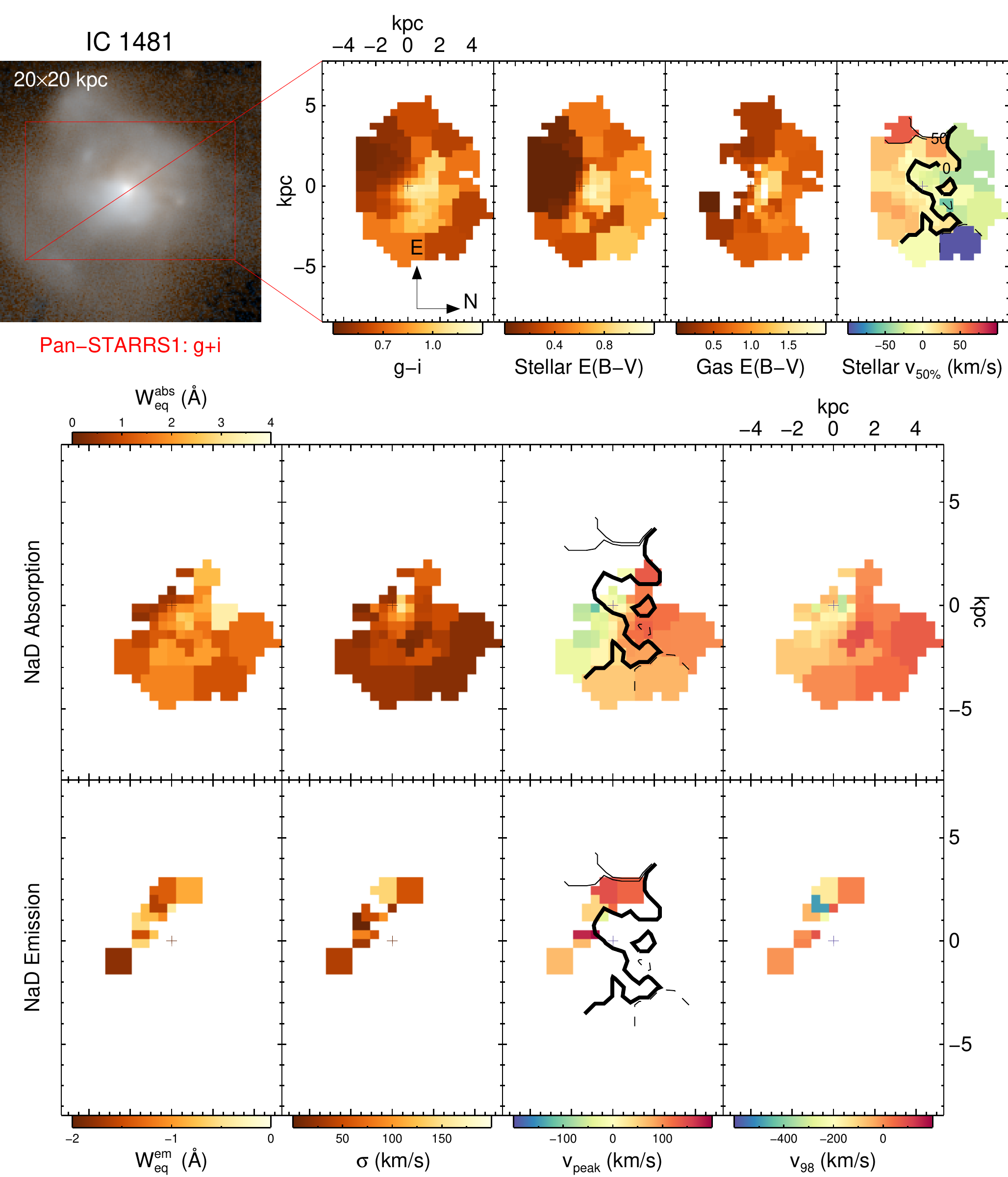}
 \caption{\it Continued.}
\end{figure*}

{\bf Figure~A1}. \nad\ absorption equivalent width vs. colour excess of the ionized gas in each galaxy, compared to MW and other external galaxy sightlines. The lines are as described in Fig. \ref{fig:weq_abs_v_ebv_gas_other}. Error bars are 1$\sigma$.\\

\begin{figure*}
 \includegraphics[width=\textwidth]{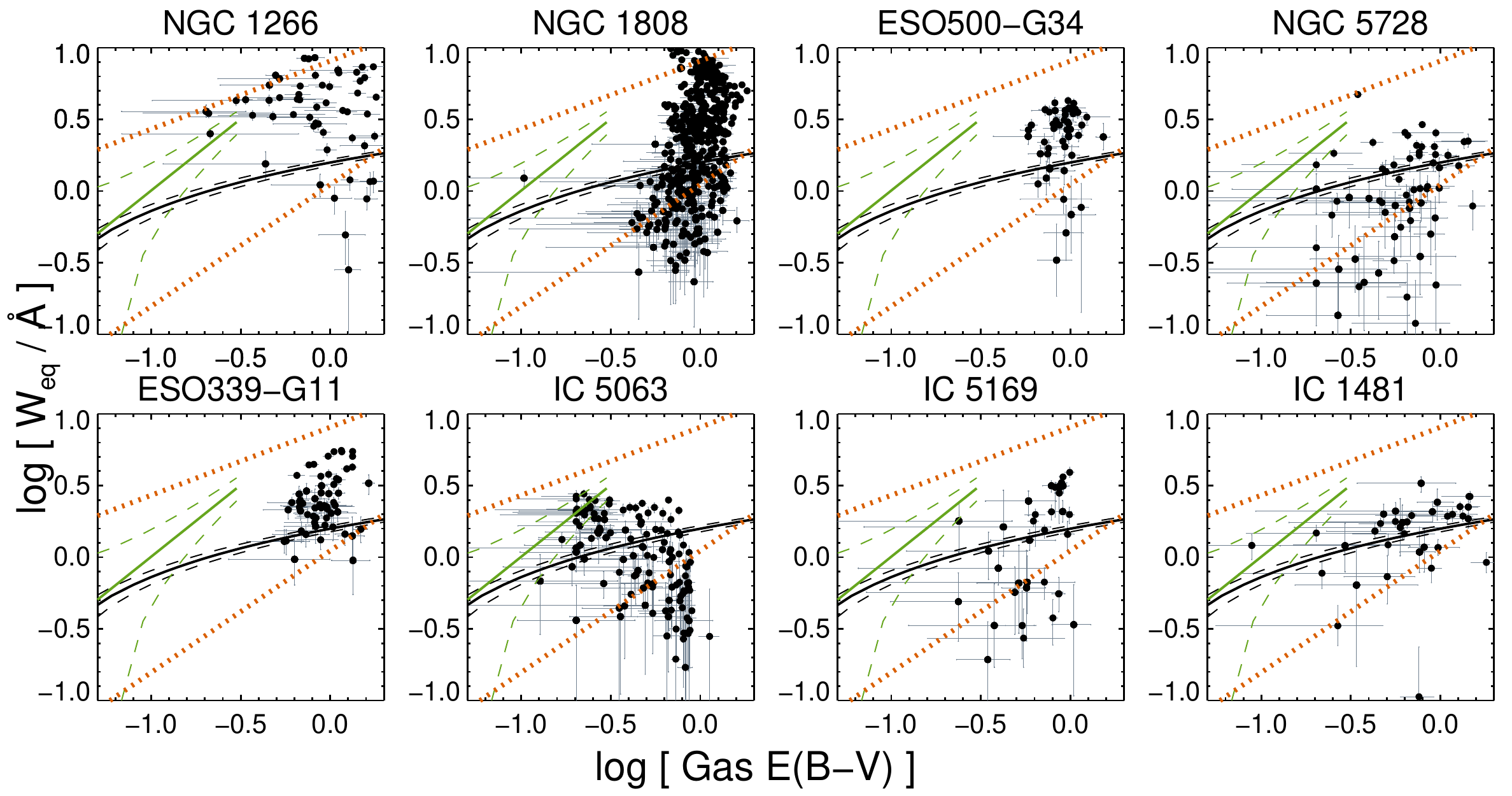}
 \label{fig:weq_abs_v_ebv_gas_bygal}
\end{figure*}

{\bf Figure~A2}. \nad\ absorption column density vs. colour excess of the ionized gas in each galaxy, compared to MW and other external galaxy sightlines. The lines are as described in Fig. \ref{fig:nnai_abs_v_ebv_gas_other}. Black filled (red open) circles represent two-component (one-component) fits. Lower limits are shown as small arrows.\\

\begin{figure*}
 \includegraphics[width=\textwidth]{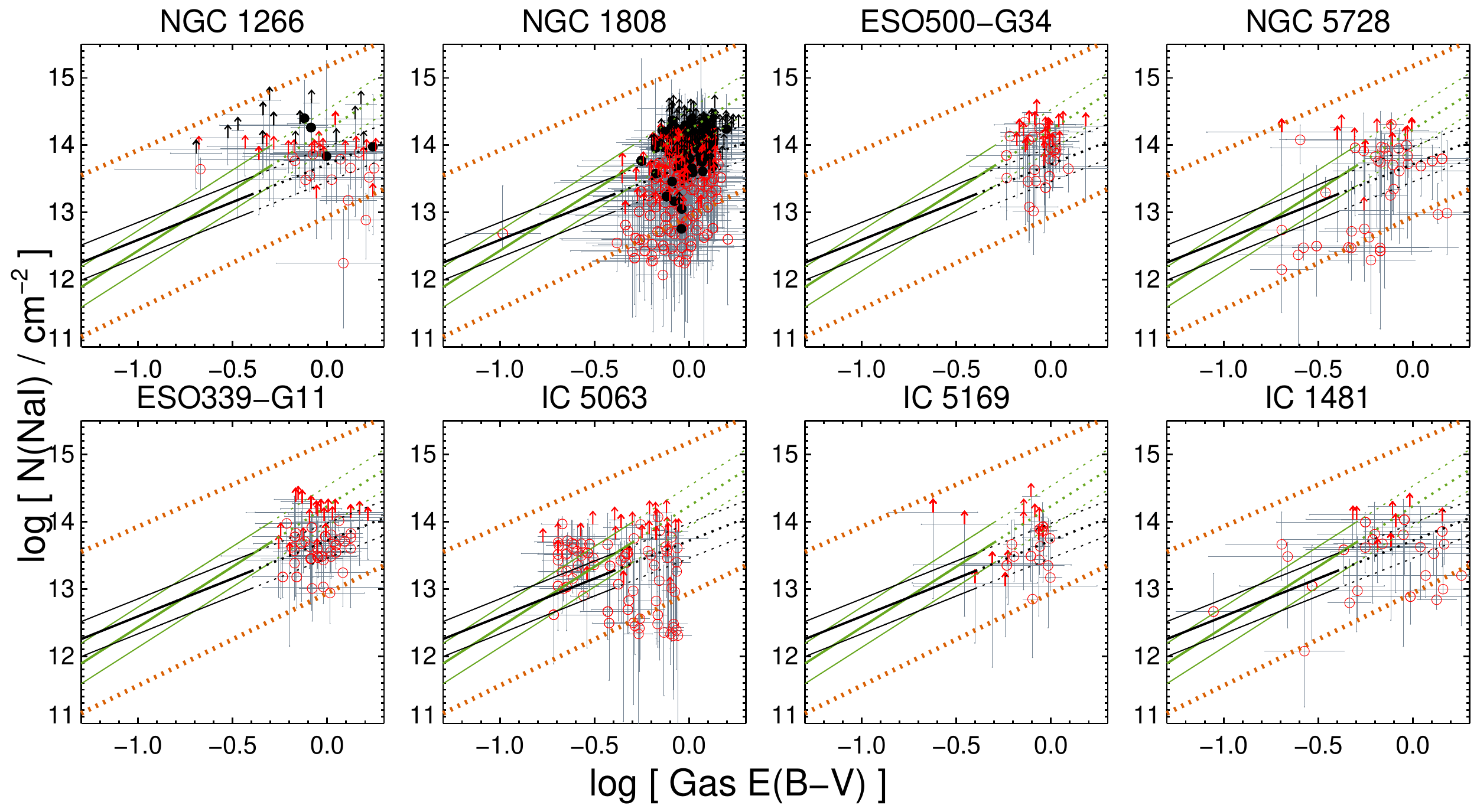}
 \label{fig:nnai_abs_v_ebv_gas_bygal}
\end{figure*}

Please note: Oxford University Press are not responsible for the content or functionality of any supporting materials supplied by the authors. Any queries (other than missing material) should be directed to the corresponding author for the paper.


\appendix

\section{Column density and equivalent width vs. colour excess for individual galaxies}
\label{appendixa}

In Section~\ref{sec:results}, we combine measurements from individual
galaxies to study $N$(\ion{Na}{1}) and \weqa\ vs. gas \ebv. Plots for
individual galaxies are available online.


\bsp	
\label{lastpage}
\end{document}